\documentclass[preprint2,times,tighten]{aastex6}
\usepackage{amsmath,amstext}
\usepackage[figure,figure*]{hypcap}
\usepackage{newtxmath} %use times font for math

\shorttitle{Probabilistic Cataloging in Crowded Fields}
\shortauthors{Portillo et al.}

\begin{document}

\title{Improved Point Source Detection in Crowded Fields using Probabilistic Cataloging}
\author{Stephen K. N. Portillo\altaffilmark{1}, Benjamin C. G. Lee\altaffilmark{2}, Tansu Daylan\altaffilmark{3}, and Douglas P. Finkbeiner\altaffilmark{1,3}}
\email{sportillo@cfa.harvard.edu}
\altaffiltext{1}{Harvard-Smithsonian Center for Astrophysics, Cambridge, MA}
\altaffiltext{2}{Harvard College, Cambridge, MA}
\altaffiltext{3}{Department of Physics, Harvard University, Cambridge, MA}

\begin{abstract}
Cataloging is challenging in crowded fields because sources are extremely covariant with their neighbors and blending makes even the number of sources ambiguous. We present the first optical probabilistic catalog, cataloging a crowded ($\sim0.1$ sources per pixel brighter than 22\textsuperscript{nd} magnitude in F606W) Sloan Digital Sky Survey r band image from M2. Probabilistic cataloging returns an ensemble of catalogs inferred from the image and thus can capture source-source covariance and deblending ambiguities. By comparing to a traditional catalog of the same image and a Hubble Space Telescope catalog of the same region, we show that our catalog ensemble better recovers sources from the image. It goes more than a magnitude deeper than the traditional catalog while having a lower false discovery rate brighter than 20\textsuperscript{th} magnitude. We also present an algorithm for reducing this catalog ensemble to a condensed catalog that is similar to a traditional catalog, except it explicitly marginalizes over source-source covariances and nuisance parameters. We show that this condensed catalog has a similar completeness and false discovery rate to the catalog ensemble. Future telescopes will be more sensitive, and thus more of their images will be crowded. Probabilistic cataloging performs better than existing software in crowded fields and so should be considered when creating photometric pipelines in the Large Synoptic \replaced{Space}{Survey} Telescope era.
\end{abstract}


\section{Introduction}
While telescopes measure the incident light from the sky and generate images, observational astronomy is often done not with those images but with catalogs constructed from those images. In cataloging, emission sources are identified and their properties (e.g. flux, color, spatial extent) are measured. These catalogs can be used to study the population of sources itself: for example, the age and metallicity of a globular cluster can be measured from a color-magnitude diagram of its stars. The sources may also be used as markers or beacons, as in the Argonaut map \citep{2015ApJ...810...25G} which uses the flux and color of stars to map the 3D distribution of the intervening dust reddening and dimming the stars. Source catalogs are also used by scientists interested in only a certain class of source to identify candidates with specific properties for further followup.

When the sources are bright and well-separated, the catalog contains the information in the original image, and interpretation of uncertainty estimates is straightforward.  However, in crowded fields, deblending becomes harder, and traditional catalogs cannot capture all of the information in the original image. For example, consider an emission peak that is best explained by two dim sources but is still consistent with a single bright source. A traditional catalog only allows a single enumeration of the sources in the image and thus must choose only one of these possibilities. This deblending ambiguity is not passed on to the downstream analyses that use the catalog, possibly causing these analyses to underestimate their uncertainties. Ambiguous sources may be flagged and cut by these analyses, but this cut potentially throws away information contained in the image. Furthermore, the inferred position and flux of a source will affect those inferred for its close neighbors. This covariance cannot be captured by a traditional catalog that simply lists the position, with uncertainty, for each source.

In addition, a consequence of the improved depth of future telescopes, such as the Large Synoptic \replaced{Space}{Survey} Telescope (LSST), is that most images will be in the crowded-field limit. Thus, current photometry algorithms will no longer suffice. For instance, in discussing potential photometry algorithms  for the LSST pipeline, \cite{2007PASP..119.1462B} note that ``\textit{no} algorithms are able to meet the \deleted{SRD }[Science Requirement Document] in \deleted{PSF }[Point Spread Function] photometry -- the numbers consistently fall short by a factor of 2-3."  Consequently, alternative photometry algorithms for cataloging crowded stellar fields and crowded fields in general are crucial to future surveys and to  astrophysical research as a whole.  

We propose probabilistic cataloging to improve the fidelity of catalogs in crowded fields. Instead of deriving a single catalog from an image, probabilistic cataloging infers the posterior distribution of catalogs in a Bayesian framework. This posterior distribution is sampled repeatedly (each sample being a catalog) to yield a catalog ensemble. Each catalog in the ensemble is a fair draw from the posterior distribution and is thus consistent with the data, and the catalog ensemble represents the state of knowledge about what sources could be in the image. In the case of ambiguous deblending detailed previously, most catalogs in the ensemble would explain the peak of emission with two dim sources but some catalogs would contain just a single bright source, reflecting the fact that a single source is still an allowed possibility. The relative prevalence of the two-source and one-source explanations in the ensemble would reflect the statistical favorability of these explanations relative to each other. The derived catalog ensemble reflects this ambiguity, allowing \deleted{straightforward }propagation of deblending uncertainties to downstream analyses. Source-source covariance is naturally captured by a catalog ensemble, and resulting uncertainties are naturally marginalized over the effects of neighbors, including neighbors too faint to appear in a traditional catalog.  This marginalization over faint sources is a non-trivial benefit.  While not all users of the catalog will need a list of possible 4-sigma sources, many users will be interested in 20-sigma sources that have errors correctly marginalized over the effects of neighboring 4-sigma sources.

Several recent efforts have begun to implement probabilistic cataloging in astronomy. \cite{2003MNRAS.338..765H} present it as ``simultaneous detection of all objects'' and apply it to a toy problem: a $200\times200$ pixel image with 8 discrete 2D Gaussian shaped objects of varying flux and radius. \cite{2013AJ....146....7B} extend its application to a crowded field, applying it to a simulated $100\times100$ optical image with 1,000 stars and showing that it outperforms SExtractor \citep{1996A&AS..117..393B}. \cite{2015ApJ...808..137J} implement probabilistic cataloging on a $25"\times25"$ \textit{Chandra} field with 14 sources, incorporating spectral models for the sources in order to use photon energy information to help deblend sources.

Other Bayesian source detection algorithms have been implemented; unlike probabilistic cataloging, they do not produce a catalog ensemble. For example, \cite{2003MNRAS.338..765H} present an iterative object detection method which stops when the Bayesian evidence disfavors adding additional sources. \cite{2009MNRAS.393..681C} speed up this approach by a factor of $\sim 100$, and further refine it in \cite{2012MNRAS.427.1384C} to produce the \textit{Planck} Early Release Compact Source Catalogue. These implementations are considerably faster than probabilistic cataloging and work well on uncrowded fields. However, these methods do not capture source-source covariances and make approximations that do not hold in crowded fields. \cite{2012MNRAS.422.1674M} offer a review of source detection approaches, including Bayesian approaches.

\added{Other works that focus on source characterization, rather than source detection, produce samples of the posterior distributions of source characteristics rather than point estimates. For example, Aeneas \citep{2011MNRAS.411..435B}, part of the \textit{Gaia} astrophysical parameters inference system \citep{2013A&A...559A..74B}, returns posterior samples of stellar properties and interstellar extinction given multiband photometry and parallaxes.

While exoplanets are not typically found by cataloging images, many groups approach the problem of detecting and characterizing exoplanets in radial velocity data with an approach similar to probabilistic cataloging. These groups report posterior distributions of system parameters, capturing the inferences that can be made from the data in a way that a point estimate cannot. For example, \cite{2005AJ....129.1706F} fits a two-planet model to 47 UMa and finds that the period distribution for the outer planet is bimodal, and \cite{0004-637X-631-2-1198} fits a single-planet model to HD 73526 and finds that the orbital period posterior is trimodal. \cite{2009MNRAS.394.1936B} present the software EXOFIT which can sample posterior distributions for models with one or two planets. \cite{doi:10.1093/mnras/stv199} allow the number of planets to be an unknown parameter, implementing fully probabilistic cataloging of radial velocity data.}

In this work, we present a probabilistic catalog inferred from part of a Sloan Digital Sky Survey (SDSS) \citep{2006AJ....131.2332G, 2011AJ....142...72E, 2012ApJS..203...21A} image of the globular cluster M2, the first probabilistic catalog constructed from optical data. We evaluate both our probabilistic catalog and a traditional catalog derived from the same SDSS image by comparing to a much deeper catalog from the Hubble Space Telescope (HST). We also introduce a procedure to distill the probabilistic catalog samples into a condensed catalog which is easier to use in downstream analyses. Like a traditional catalog, it lists sources with their properties and uncertainties, but unlike a traditional catalog, the condensed catalog also marginalizes the sources' properties over their neighbors and includes a measure of confidence in each source's existence.

The rest of this paper is organized as follows. In Section \ref{sec:observations}, we describe the SDSS observations used to construct our probabilistic catalog, the traditional catalog we compare against, and the HST catalog we use as ground truth to assess both catalogs. In Section \ref{sec:methods}, we detail our Bayesian point source model and the MCMC sampler that we use. In Section \ref{sec:pcat}, we present our probabilistic catalog and compare it against a traditional catalog. In Section \ref{sec:condensed}, we present our condensed catalog and compare it to the full probabilistic catalog. We also compare the reported flux errors to the flux errors expected in a sparse field in order to identify spurious sources. In Section \ref{sec:discussion}, we discuss implications and conclude.

\section{Observations}
\label{sec:observations}
In order to test the performance of probabilistic cataloging against traditional cataloging, we chose an image of the globular cluster M2 from SDSS. Globular clusters are rich with stars, providing challenging crowded fields to catalog. In addition, in globular clusters, nearly every source is a star rather than a galaxy, allowing the image to be modeled using only point sources. \citet{2008ApJS..179..326A} cataloged the same SDSS image of M2 using DAOPHOT \citep{1987PASP...99..191S, 1994chst.conf...89S}, providing a traditional catalog (\replaced{denoted}{referred to hereafter as} the ``DAOPHOT catalog''\deleted{ in in what follows}) to which we can directly compare our probabilistic catalog. In addition, the ACS Globular Cluster survey \citep{2007AJ....133.1658S} cataloged M2 using imaging from the Advanced Camera for Surveys on HST. HST has $\approx 20$ times the angular resolution of the Sloan Foundation 2.5 m Telescope, so the stars are much more widely separated in the HST image and are thus easily picked up by traditional cataloging methods. In addition, the HST image has $\approx 30$ times the exposure of the SDSS image (five 340 s HST exposures vs the 34.9 s SDSS exposure). We take the catalog made from the HST data (\replaced{denoted}{referred to hereafter as} the ``HST catalog''\deleted{ in what follows}) as a reliable reference against which catalogs made from the SDSS data can be judged.

M2 is a rich and compact globular cluster containing about 150,000 stars located 11.5 kpc away. It has a core radius of 0.34 arcminutes (1.1 parsecs) and a half-light radius of 1.08 arcminutes (3.6 parsecs) \citep[2010 edition]{1996AJ....112.1487H}.

Our main test image is SDSS run 2583, field 136, camcol 2, in the r band.  There is no SDSS catalog for this field, because the survey photometric pipeline, Photo \citep{2001ASPC..238..269L}, timed out.  For most of our tests, we focus on a $100 \times 100$ pixel cutout from the image $\approx 2$ arcminutes from the center of the cluster.  This patch is crowded, but away from the heavily saturated core.  The HST catalog identifies 1,049 sources brighter than 22\textsuperscript{nd} magnitude in the F606W band in this patch. F606W is much broader than SDSS r, but centered at roughly the same wavelength, and F606W magnitudes for main sequence stars agree with SDSS r magnitudes to a few tenths of a magnitude.

\section{Methods}
\label{sec:methods}
\subsection{Model Space and Priors}
The model space is the space of possible point-source catalogs. We assume that there are an unknown number, $N$, of point sources in the image. In our case of point sources in a single band, each source $i$ is described by its position $(x_i,y_i)$ and flux $f_i$. We assume the sources are uniformly distributed spatially and that their fluxes follow a power-law distribution \added{with index 2} between $f_{min}$, which is fixed, and $f_{max}$, which is unknown. The reciprocal of $f_{max}$ is given a uniform prior on $(0, f_{bound}^{-1})$, where $f_{bound}$ is slightly dimmer than the brightest point source in the image, forcing $f_{max}$ to be at least as bright as the brightest point source.\deleted{ The index $\alpha$ of the power-law is also taken to be unknown. The reciprocal of $\alpha$ is given a uniform prior on $(0,1)$ so that the flux distribution falls as at least $f^{-1}$.}

In addition to the point source emission, the image has unknown uniform sky emission $I_{sky}$, which is given a log-uniform prior. Thus, a single catalog can be described by a parameter vector $\theta$: \deleted{$\alpha$ in Equation \ref{eqn:catalogdef}}
\begin{equation}
\label{eqn:catalogdef}
\theta = \{N, \{x_i,y_i,f_i\}_{i=1}^N , f_{max}, I_{sky}\}.
\end{equation}
Note that the catalog, as defined above, includes not only the point sources and their properties (which would constitute a \replaced{"catalog"}{``catalog''} in normal usage), but also the hyperparameters describing the point source population and the sky level as a nuisance parameter. Also, a traditional catalog would include uncertainties on the positions and fluxes, but each sample of a probabilistic catalog does not have uncertainties on its parameters. Instead, the uncertainty manifests in the distribution of these parameters across the catalog ensemble.

The space of possible catalogs is transdimensional, that is to say, not of fixed dimension. With 3 parameters for each source and \replaced{3 parameters describing the flux distribution of sources}{2 other parameters}, the space of possible catalogs with $N$ sources is \replaced{$3N+3$}{$3N+2$} dimensional. The space of all possible catalogs can thus be thought of as the union of the spaces of possible catalogs with $N=0,1,2,3$, and so on. It thus contains fixed-dimensional subspaces each with a different dimensionality. Meaningful priors can still be assigned on this space while the likelihood is in fixed-dimension data space, so a posterior over the space of possible catalogs is definable. Such a transdimensional space can be sampled with methods like Reversible Jump MCMC \citep{GREEN01121995} or Birth Death MCMC \citep{stephens2000}.

We treat the number of point sources, $N$, as arising from a Poisson process which would yield $\mu$ point sources on average in the imaged region. \citet{2016arXiv160704637D} is concerned with constraining populations of sub-threshold sources and treats $N$ as the answer to the question: ``What is the number of point sources above a given minimum flux, $f_{min}$ (counting even sub-threshold sources)?" Thus, they place a log uniform prior on $\mu$, not wanting to penalize possible sub-threshold sources. However, the current work focuses on deblending overlapping and significant sources. We are not interested in the population of sub-threshold sources, and they will not greatly affect the deblending of highly significant sources. Since the time taken to evaluate the likelihood for each sample increases with the number of included point sources, we instead \replaced{use a prior on $\mu$}{add a prior on $N$} that penalizes additional point sources and cuts out these sub-threshold sources. For a Gaussian problem, when all parameters are away from the boundaries and the data are not overfit, adding a source improves the log likelihood of the maximum likelihood solution, on average, by $\frac{3}{2}$ ($\frac{1}{2}$ for each of three parameters $x_i,y_i,f_i$). Thus we choose a prior that counteracts the expected log likelihood gain from additional sources: \explain{$\mu$ replaced with $N$ in Equation \ref{eqn:prior_nplusone}}
\begin{equation}
\label{eqn:prior_nplusone}
\log \frac{\pi(N+1)}{\pi(N)} = -\frac{3}{2}.
\end{equation}
This choice implies an exponential prior on \replaced{$\mu$}{$N$}: \explain{$\mu$ replaced with $N$ in Equation \ref{eqn:exponential_nprior}}
\begin{equation}
\label{eqn:exponential_nprior}
\pi(N) \propto \exp\left(-\frac{3}{2} N\right).
\end{equation}

We set $f_{min}$ to correspond to the SDSS 95\% completeness limit in a sparse field ($\approx 22$ mag in r band), expecting the dimmest significant source in the crowded field to be brighter than this limit.

\added{
Our model space and priors are very similar to those in \cite{2013AJ....146....7B}, except for a few differences. We put a more restrictive prior on $N$ and we don't use a broken power-law flux distribution. Most significantly, we do not fit for the point spread function (PSF), but take it as known.
}

\subsection{Generative Model and Likelihood}
To calculate the likelihood of a given catalog, the corresponding model image must be produced. The data are photoelectron counts $k_{lm}$ in a $W \times H$ grid of pixels with coordinates $(x,y)=(l,m)$. We use the pixel-convolved point spread function extracted by the standard SDSS pipeline for the center of our image, $\mathcal{P}(\Delta x, \Delta y)$, to predict the expected counts $\lambda_{lm}$ for each pixel from a uniform sky background $I_{sky}$ and nearby sources:
\begin{equation}
\label{eqn:pointsourcesum}
\lambda_{lm} = I_{sky} + \sum_{i=1}^{N} f_i \mathcal{P}(l - x_i, m - y_i).
\end{equation}
In the limit of many photoelectron counts, the noise when $\lambda_{lm}$ photoelectrons are expected is Gaussian with standard deviation $\sqrt{\lambda_{lm}}$ photoelectrons. The likelihood is then
\begin{equation}
\label{eqn:likelihood}
\mathcal{L} = \prod_{l=1}^{W} \prod_{m=1}^{H} \frac{1}{\sqrt{2\pi\lambda_{lm}}} \exp\left(-\frac{(k_{lm}-\lambda_{lm})^2}{2 \lambda_{lm}}\right).
\end{equation}
Taking the log likelihood and dropping additive constants because MCMC sampling only requires likelihood ratios, $\log \mathcal{L}$ gives:
\begin{equation}
\label{eqn:loglikelihood}
\log \mathcal{L} = \sum_{l=1}^{W} \sum_{m=1}^{H} -\frac{1}{2} \log \lambda_{lm} - \frac{(k_{lm}-\lambda_{lm})^2}{2 \lambda_{lm}}.
\end{equation}
The first term, being a logarithm, is much less sensitive to changes in the $\lambda_{ij}$ than the second term. Assuming the first term is constant reduces the log likelihood evaluations to only additions and multiplications, greatly speeding up sampling:
\begin{equation}
\label{eqn:simpleloglike}
\log \mathcal{L} \approx \sum_{l=1}^{W} \sum_{m=1}^{H} - \frac{(k_{lm}-\lambda_{lm})^2}{2 \lambda_{lm}}.
\end{equation}

\subsection{MCMC Sampling}
This work performs MCMC sampling of the posterior distribution using Diffusive Nested Sampling\footnote{We use the DNest3 C++ implementation, available at \url{https://github.com/eggplantbren/DNest3}.} \citep{Brewer2011}. Rather than directly sampling the posterior distribution as in Metropolis-Hastings, Diffusive Nested Sampling constructs an alternative target distribution. The Diffusive Nested Sampling algorithm first samples the prior and then uses these samples to determine a likelihood threshold that contains $\approx e^{-1}$ of the prior mass above it. The sampler then samples from a modified version of the prior where the parameter space above the likelihood threshold (the first level) is upweighted. Using these samples, a second likelihood threshold is determined such that the parameter space above it (the second level) contains $\approx e^{-2}$ of the prior mass above it. The sampler then samples from a modified version of the prior where the first level is still upweighted, but the second level is upweighted more. This process continues until the next level constructed is determined to hold so little of the posterior mass that it is unimportant.

In constructing these levels, the sampler behaves similarly to simulated annealing, but with a cooling schedule tailored to the posterior distribution using samples. Once enough levels have been constructed, the sampler continues to sample from this series of levels. The resulting samples can later be reweighted to yield samples from the posterior. This target distribution allows the sampler to escape posterior peaks by sampling from a lower (and thus less constraining) level; the sampler is even allowed to return to the prior. This freedom allows the sampler to avoid being trapped in a single posterior peak and instead find multiple peaks in a multi-modal distribution.

\section{Probabilistic Catalog}
\label{sec:pcat}
\subsection{Computational Requirements}
All code was run on a machine with two six-core Intel Xeon E5-2667 processors at 2.9 GHz. Running our probabilistic cataloging code using all 12 cores on the selected $100 \times 100$ pixel image, inferring $\approx 1,100$ sources, required a day of wall clock time to yield useful results. Each sample in the MCMC chain requires 1 core-ms on average.

\subsection{Representation as Catalog Ensemble}

\begin{figure*}
\plotone{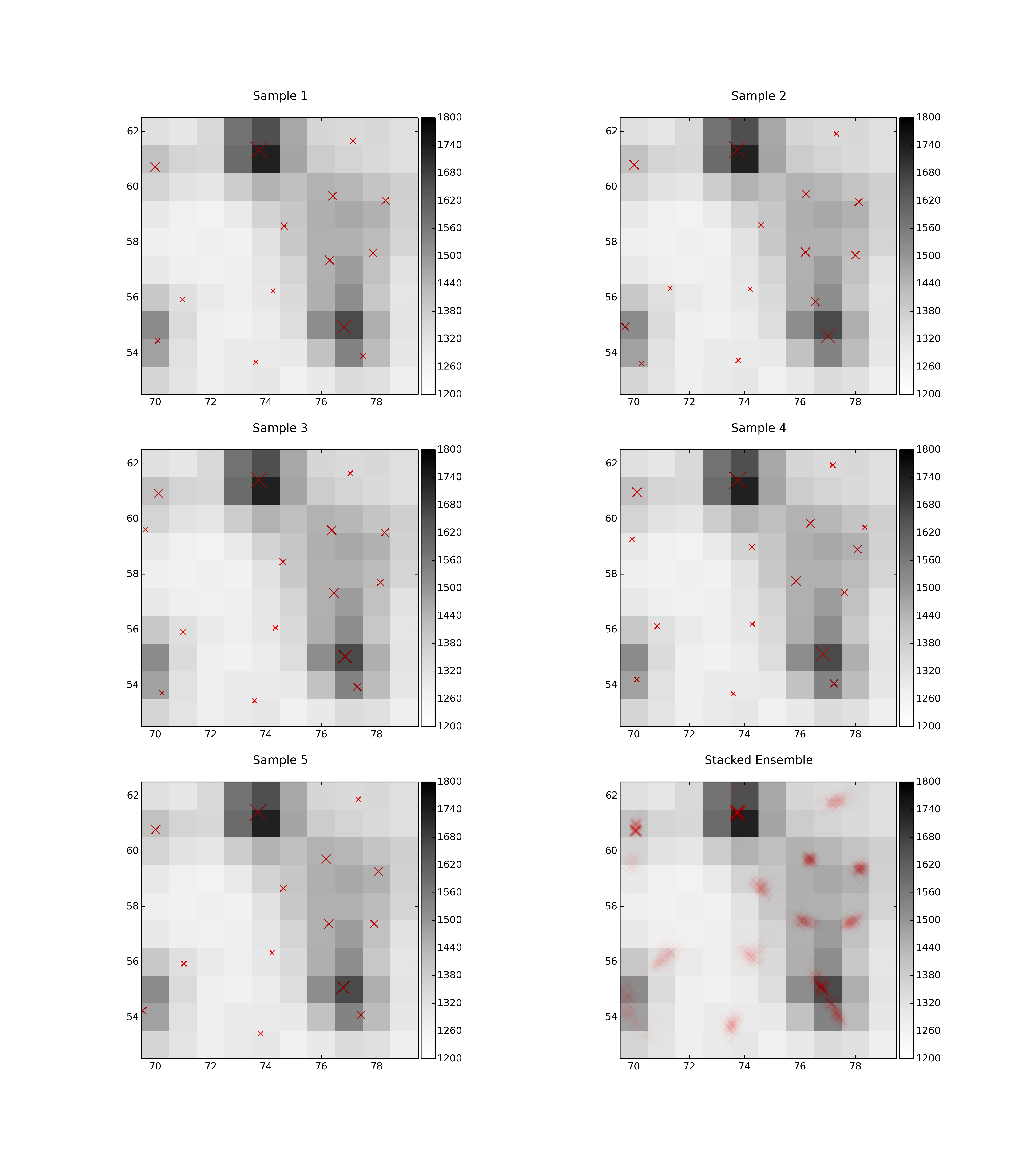}
\caption{Depiction of the catalog ensemble in a 10x10 pixel cutout of the SDSS image. Five fair samples from the catalog ensemble are plotted with red Xs with the symbol area proportional to the source flux. Many fair samples with translucent red Xs are plotted in the sixth panel, showing how errors in position are captured by the catalog ensemble.}   \label{fig:samples}
\end{figure*}

To clarify the representation of the probabilistic catalog as a catalog ensemble, we present fair samples from the ensemble in Figure \ref{fig:samples}. Sources that are inferred confidently appear in all of the sample catalogs, while less confident inferences only appear in a fraction of the samples. The uncertainty in the position and flux of a confidently identified source manifests as the distribution of these properties across sample catalogs. The uncertainty in deblending ambiguous sources is reflected in the differing deblendings made in sample catalogs. \deleted{As with any Markov chain, these uncertainties may be propagated through a downstream analysis simply by repeating the analysis once for each posterior sample, yielding draws from the posterior pdf of the result. The transdimensionality adds no new complications to the use of the catalog ensemble, although increased computation time may be burdensome.}

\added{To correctly use the catalog ensemble in a further hierarchical inference, the samples need to be decorated with their prior values. Then, using the approach presented in \cite{0004-637X-725-2-2166} and \cite{0004-637X-795-1-64}, if a downstream user wishes to use the catalog ensemble to infer some hyperparameters $\eta$ (say, the core radius of the cluster), the likelihood in $\eta$ can be calculated using samples from the catalog ensemble (up to a multiplicative constant):
\begin{align}
\label{eqn:hierarchical_like}
\mathcal{L}(\{k_{lm}\}|\eta) &= \int \frac{\pi(\theta|\eta)}{\pi(\theta|\alpha)} P(\theta|\{k_{lm}\}, \alpha) d\theta \\
\label{eqn:hierarchical_like_2}
&\approx \sum_{\theta \sim P(\theta|\{k_{lm}\}, \alpha)} \frac{\pi(\theta|\eta)}{\pi(\theta|\alpha)},
\end{align}
where the notation $\alpha$ is used as a reminder that the priors used in constructing the catalog ensemble, $\pi(\theta|\alpha)$, may be different than those desired in the hierarchical inference, $\pi(\theta|\eta)$. Clearly, if the desired priors for the hierarchical inference agree with those used to produce the catalog ensemble, then the catalog samples can be used directly as the ratio in Equation \ref{eqn:hierarchical_like_2} is exactly unity. Expanding out the priors used in this work gives:
\begin{align}
\pi(\theta|\alpha) = &\pi(N|\alpha)\pi(f_{max}|\alpha)\pi(I_{sky}|\alpha) \\ \nonumber
\label{eqn:full_prior}
&\prod_{i=1}^{N} \pi(x_i, y_i|\alpha) \pi(f_i|f_{max}, N, \alpha).
\end{align}
The model used in this work only has three full-catalog parameters (as opposed to the $3N$ source parameters): $f_{max}$, $I_{sky}$, and N. $f_{max}$ does not affect the flux priors $\pi(f_i|f_{max}, N, \alpha)$ very much, and it is unlikely that further inferences about the characteristics of the sources will demand a specific prior $\pi(I_{sky}|\eta) \neq \pi(I_{sky}|\alpha)$. However, the prior on $N$ and the choice of an index 2 power-law flux distribution (one of the assumptions reflected by $\alpha$) prove to be more problematic, as they affect the inclusion of fainter sources. Thus, the catalog ensemble may not adequately cover the part of catalog space that would be preferred by the subsequent analysis (eg. the current catalog ensemble does not have any catalogs with flux distributions that look like power-laws of index different than 2 or like broken power-laws). If a further inference is desired which depends on the properties of fainter sources, a catalog ensemble should be constructed using the priors appropriate for that analysis.

We also argue that if a further inference mainly depends on bright sources, the catalog ensemble is useful without having to re-create it with a different prior. For example, one could measure the core radius of the cluster, but using only stars that are bright enough that the choice of priors don't affect them. Then Equation \ref{eqn:hierarchical_like_2} can be used, as the catalog ensemble will adequately cover the bright sources demanded by the measurement of the core radius. } 

\subsection{Completeness}
\label{sec:completeness}
We take the HST catalog as having the true identities and positions of the stars in the selected patch of M2. Because of HST's superior angular resolution, the HST image is much less crowded (in terms of sources per resolution element) than the SDSS image, so identifying stars is straightforward, and their positions are measured well.  Although the formal error on the HST positions is negligible, we are still limited by how well the HST and SDSS astrometric solutions are matched, and there may be noticeable movement of high proper-motion stars.

\begin{figure}
\plotone{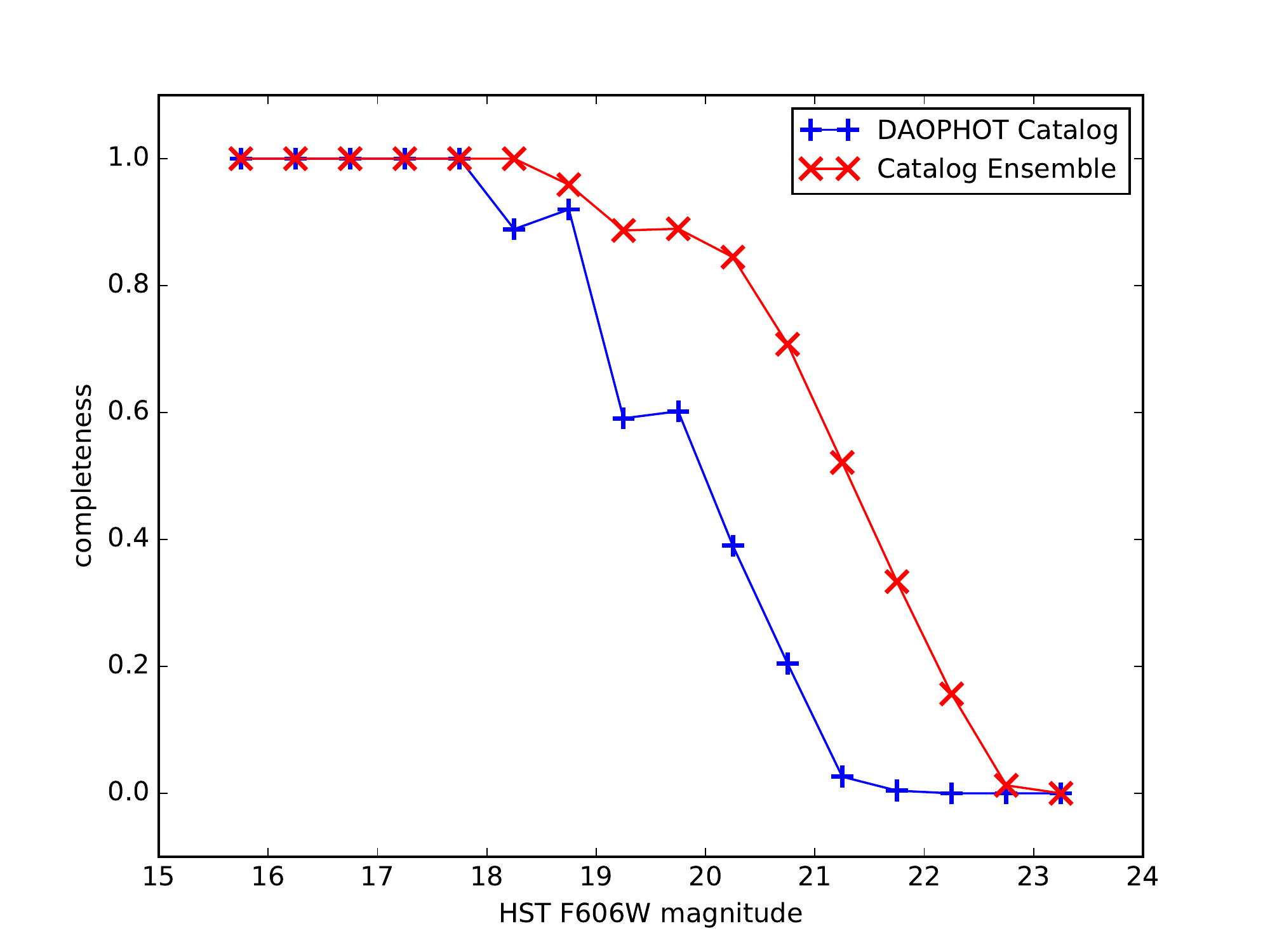}
\caption{Completeness of the catalog ensemble and the DAOPHOT catalog. Completeness is determined by comparing to the HST catalog.}
    \label{fig:completeness}
\end{figure}

In Figure \ref{fig:completeness}, we present a comparison of the completeness of our catalog ensemble and the DAOPHOT catalog against the stars in the HST catalog. Completeness is plotted against HST 606W magnitudes, the closest HST band available in this field to SDSS r band. A source is taken to be a match if it is within 0.75 pixels of the HST position (cp. the full width at half maximum of the SDSS point spread function, 2.2 pixels) and its r magnitude is within 0.5 mag of the 606W magnitude. This matching criterion is loose for two reasons: first, 606W is a wider band than r; and second, some stars have faint neighbors that are detectable in the HST image but are completely blended in the SDSS image. Because the different samples of the catalog ensemble can contain differing sources, the catalog ensemble completeness is reported as the average of the completeness of the sample catalogs. If the ensemble has $M$ samples, this averaging is equivalent to counting a match in a single sample as $M^{-1}$ of a match. The DAOPHOT catalog starts losing completeness at 18\textsuperscript{th} magnitude, becoming incomplete by 22\textsuperscript{nd} magnitude. The catalog ensemble, however, goes over 1 magnitude deeper at every level of completeness.

\subsection{False Discovery Rate}

A catalog with higher completeness is not necessarily a better catalog unless it also has a comparable or lower false discovery rate. Phrased differently, a catalog with sources at all possible positions and fluxes trivially has 100\% completeness, but is a useless catalog because nearly every source is a false positive. Borrowing terminology from the binary classification literature, we define the false discovery rate ($FDR$) in terms of the number of true positives ($TP$) and false positives ($FP$):

\begin{equation}
\label{eqn:fdr}
FDR = \frac{FP}{FP+TP}
\end{equation}

\deleted{, }with a true positive being a catalog source with an HST source satisfying the same match criteria as for completeness in Section \ref{sec:completeness}, and a false positive being a catalog source without any such match. In other words, the FDR is the fraction of the sources in the catalog which are false positives. Similarly to when we calculate completeness, when considering $M$ samples from the catalog ensemble, we count each source in a sample catalog as $M^{-1}$ of a (true or false) positive.

\begin{figure}
\plotone{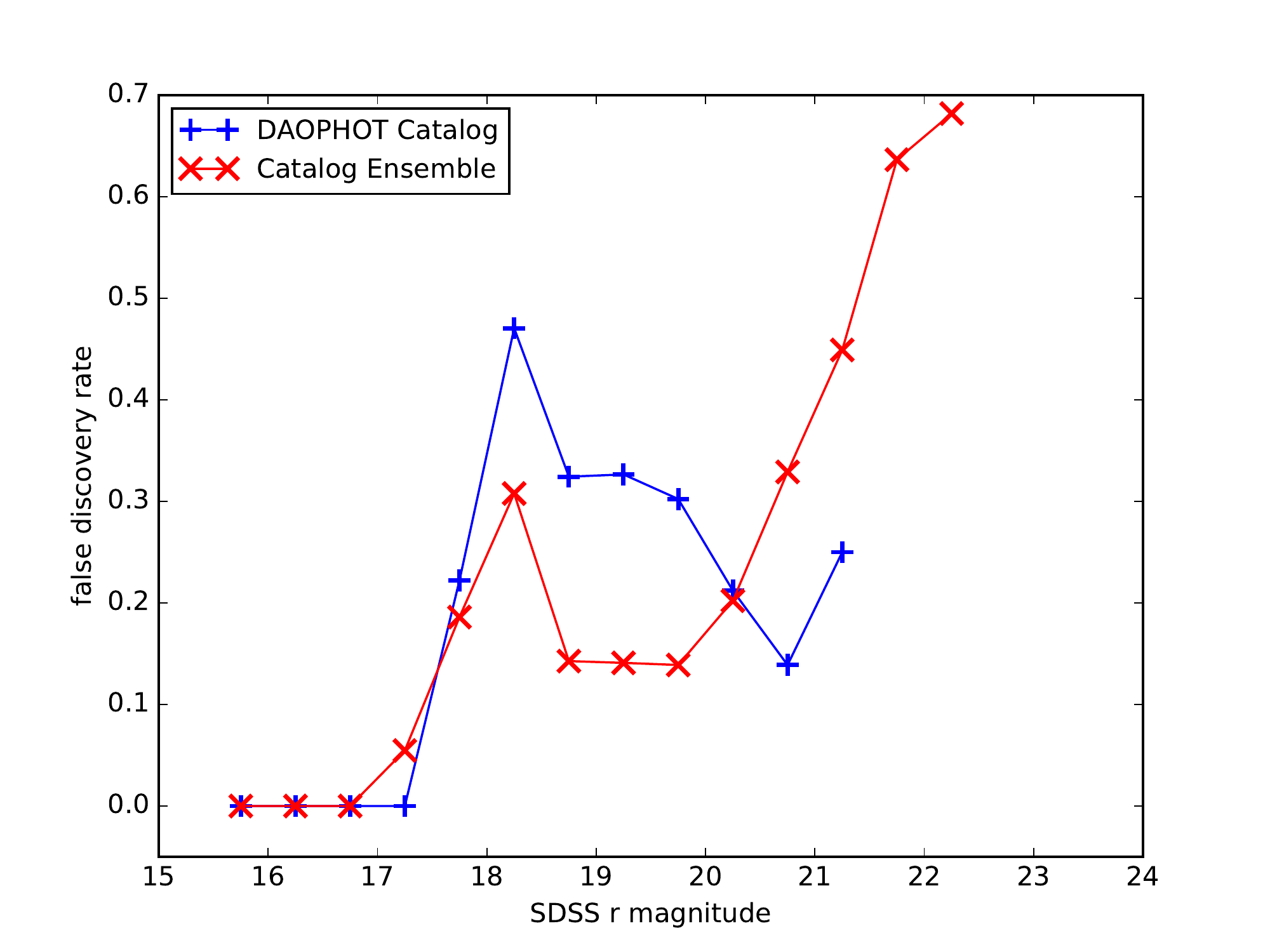}
\caption{False discovery rate of the catalog ensemble and the DAOPHOT catalog. False discovery rate is determined by comparing to the HST catalog.}
    \label{fig:falsediscovery}
\end{figure}

In Figure \ref{fig:falsediscovery}, we present a comparison of the false discovery rate of the catalog ensemble and the DAOPHOT catalog, as a function of SDSS r band magnitude. The DAOPHOT catalog false discovery rate peaks at 47\% at 18\textsuperscript{th} magnitude and stays above 10\% for fainter magnitudes. The catalog ensemble false discovery rate is lower than the DAOPHOT false discovery rate until 20\textsuperscript{th} magnitude. The false positive rate rises because closely-separated pairs of dim sources are fit with a single brighter source which does not meet the match criterion for either dim source. Beyond 21\textsuperscript{nd} magnitude, the DAOPHOT catalog cannot be assigned a false discovery rate because it does not identify any sources that faint.

\subsection{Residuals}

\begin{figure*}
\plotone{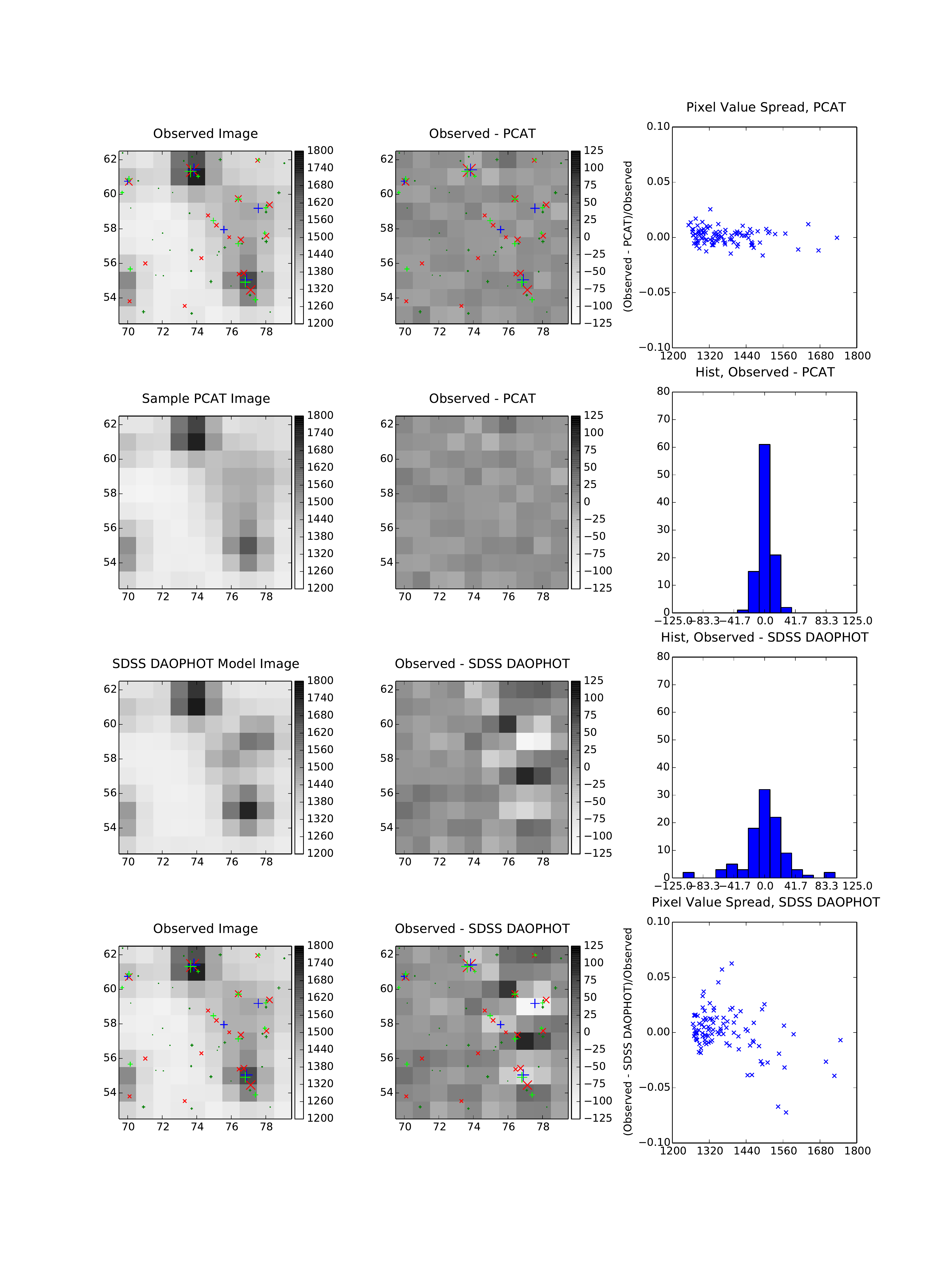}
\caption{Comparison of the source locations and residuals of the DAOPHOT catalog and catalog ensemble in a 10$\times$10 pixel cutout of the the SDSS image. Lime green crosses are HST sources brighter than 22\textsuperscript{nd} magnitude in F606W and dark green crosses are HST sources dimmer than this magnitude, both with areas proportional to F606W flux. Blue crosses are DAOPHOT sources and \deleted{translucent} red Xs are sources from \added{a fair sample of} the catalog ensemble\deleted{, stacked}. The area of the DAOPHOT catalog and catalog ensemble source symbols is proportional to SDSS r band flux. The left column shows the observed image with catalogs overplotted and the mean model images for the DAOPHOT catalog and catalog ensemble. The middle column shows the residuals of the mean model images with respect to the observed image, with and without source symbols. The right column shows the DAOPHOT and mean catalog ensemble pixel residuals histogrammed, as well as the fractional residuals scatter plotted against the observed image pixel values.}
    \label{fig:twelve_panel}
\end{figure*}

\begin{figure*}
\plotone{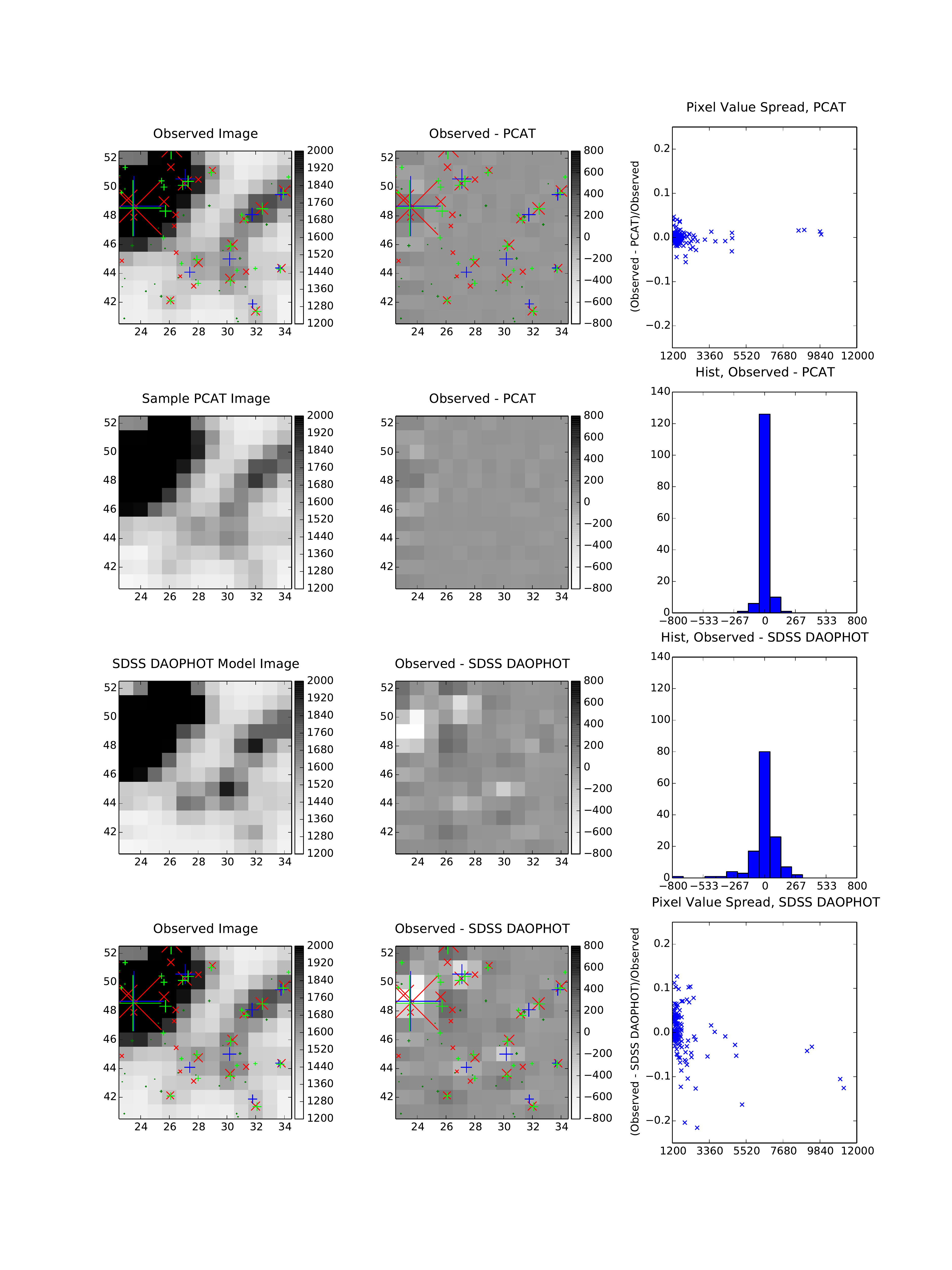}
\caption{Comparison of the source locations and residuals of the DAOPHOT catalog and catalog ensemble in a 10$\times$10 pixel cutout of the the SDSS image. Lime green crosses are HST sources brighter than 22\textsuperscript{nd} magnitude in F606W and dark green crosses are HST sources dimmer than this magnitude, both with areas proportional to F606W flux. Blue crosses are DAOPHOT sources and \deleted{translucent} red Xs are sources from \added{a fair sample of} the catalog ensemble\deleted{, stacked}. The area of the DAOPHOT catalog and catalog ensemble source symbols is proportional to SDSS r band flux. The left column shows the observed image with catalogs overplotted and the mean model images for the DAOPHOT catalog and catalog ensemble. The middle column shows the residuals of the mean model images with respect to the observed image, with and without source symbols. The right column shows the DAOPHOT and mean catalog ensemble pixel residuals histogrammed, as well as the fractional residuals scatter plotted against the observed image pixel values.}
    \label{fig:twelve_panel2}
\end{figure*}

\begin{figure*}
\plotone{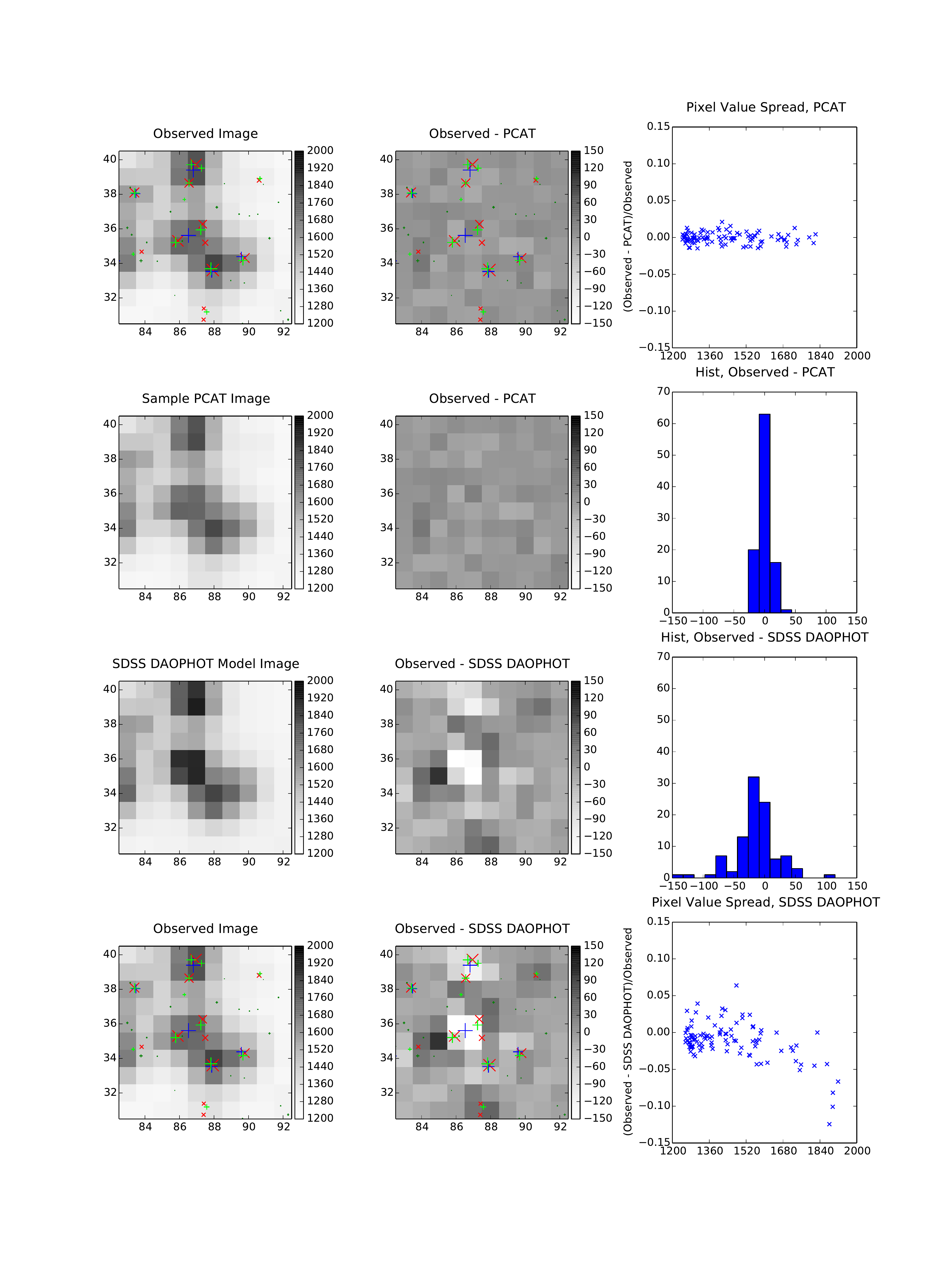}
\caption{Comparison of the source locations and residuals of the DAOPHOT catalog and catalog ensemble in a 10$\times$10 pixel cutout of the the SDSS image. Lime green crosses are HST sources brighter than 22\textsuperscript{nd} magnitude in F606W and dark green crosses are HST sources dimmer than this magnitude, both with areas proportional to F606W flux. Blue crosses are DAOPHOT sources and \deleted{translucent} red Xs are sources from \added{a fair sample of} the catalog ensemble\deleted{, stacked}. The area of the DAOPHOT catalog and catalog ensemble source symbols is proportional to SDSS r band flux. The left column shows the observed image with catalogs overplotted and the mean model images for the DAOPHOT catalog and catalog ensemble. The middle column shows the residuals of the mean model images with respect to the observed image, with and without source symbols. The right column shows the DAOPHOT and mean catalog ensemble pixel residuals histogrammed, as well as the fractional residuals scatter plotted against the observed image pixel values.}
    \label{fig:twelve_panel3}
\end{figure*}

\begin{figure*}
\plotone{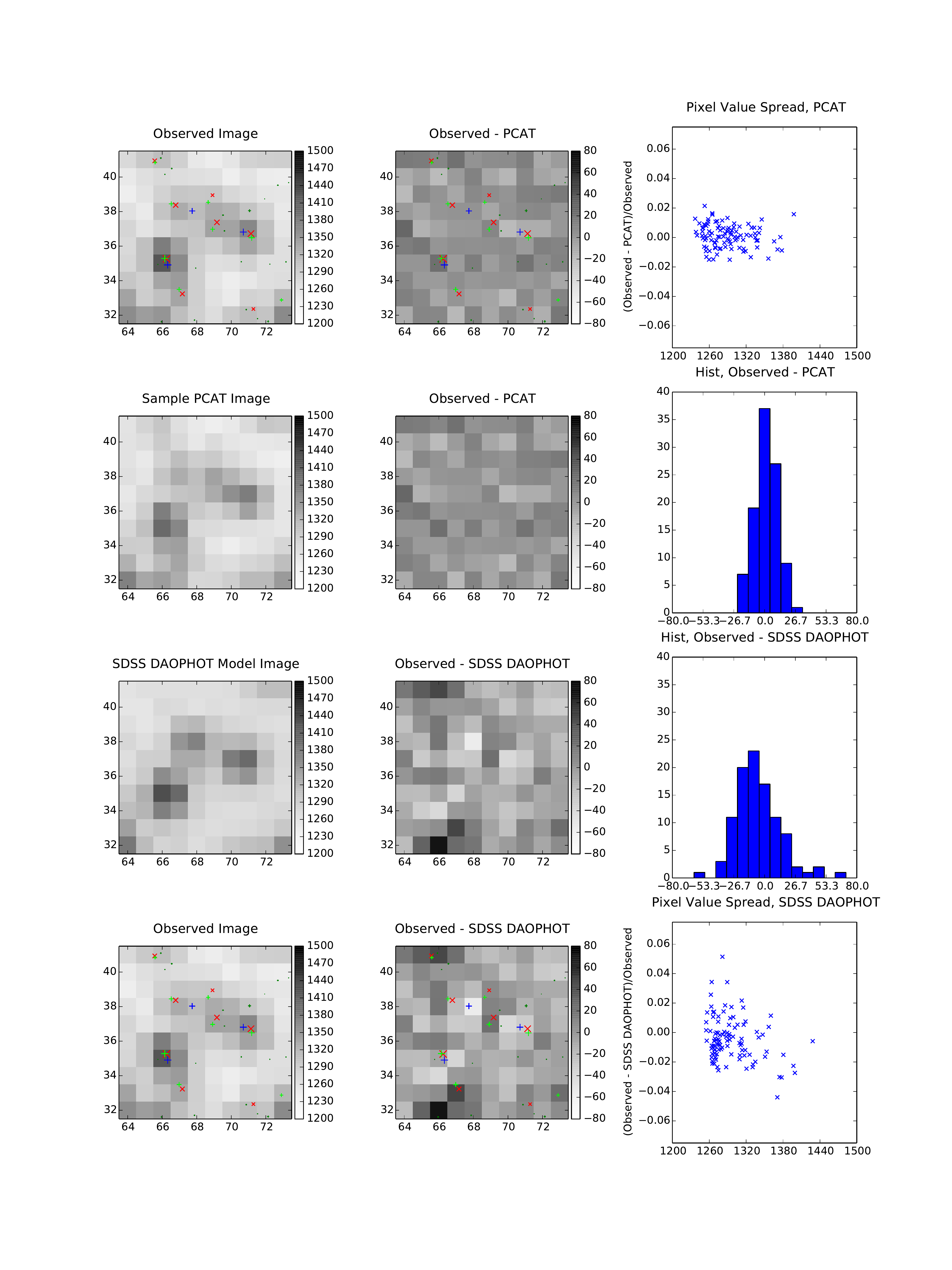}
\caption{Comparison of the source locations and residuals of the DAOPHOT catalog and catalog ensemble in a 10$\times$10 pixel cutout of the the SDSS image. Lime green crosses are HST sources brighter than 22\textsuperscript{nd} magnitude in F606W and dark green crosses are HST sources dimmer than this magnitude, both with areas proportional to F606W flux. Blue crosses are DAOPHOT sources and \deleted{translucent} red Xs are sources from \added{a fair sample of} the catalog ensemble\deleted{, stacked}. The area of the DAOPHOT catalog and catalog ensemble source symbols is proportional to SDSS r band flux. The left column shows the observed image with catalogs overplotted and the mean model images for the DAOPHOT catalog and catalog ensemble. The middle column shows the residuals of the mean model images with respect to the observed image, with and without source symbols. The right column shows the DAOPHOT and mean catalog ensemble pixel residuals histogrammed, as well as the fractional residuals scatter plotted against the observed image pixel values.}
    \label{fig:twelve_panel4}
\end{figure*}

It is instructive to look more closely at some very crowded regions of the image.  In Figures \ref{fig:twelve_panel}, \ref{fig:twelve_panel2}, \ref{fig:twelve_panel3}, and \ref{fig:twelve_panel4}, we overplot 300 stacked probabilistic catalog samples on a $10\times10$ pixel region of the SDSS image, along with the DAOPHOT catalog and the HST catalog. Consider Figure \ref{fig:twelve_panel}: the DAOPHOT catalog claims that there are five sources.  Comparing to the HST catalog, three of the DAOPHOT sources correspond to bright HST sources, while two of them correspond to the blending of several closely spaced HST sources. The probabilistic catalog captures the same three bright HST sources, better deblends the closely spaced stars into five, and places additional sources corresponding to single HST sources or collections of closely spaced HST sources. The probabilistic catalog residuals are smaller than those of the DAOPHOT catalog. Because of the probabilistic catalog's better deblending, the probabilistic catalog residuals are much smaller around crowded stars than the DAOPHOT catalog residuals.

\begin{figure*}
\plotone{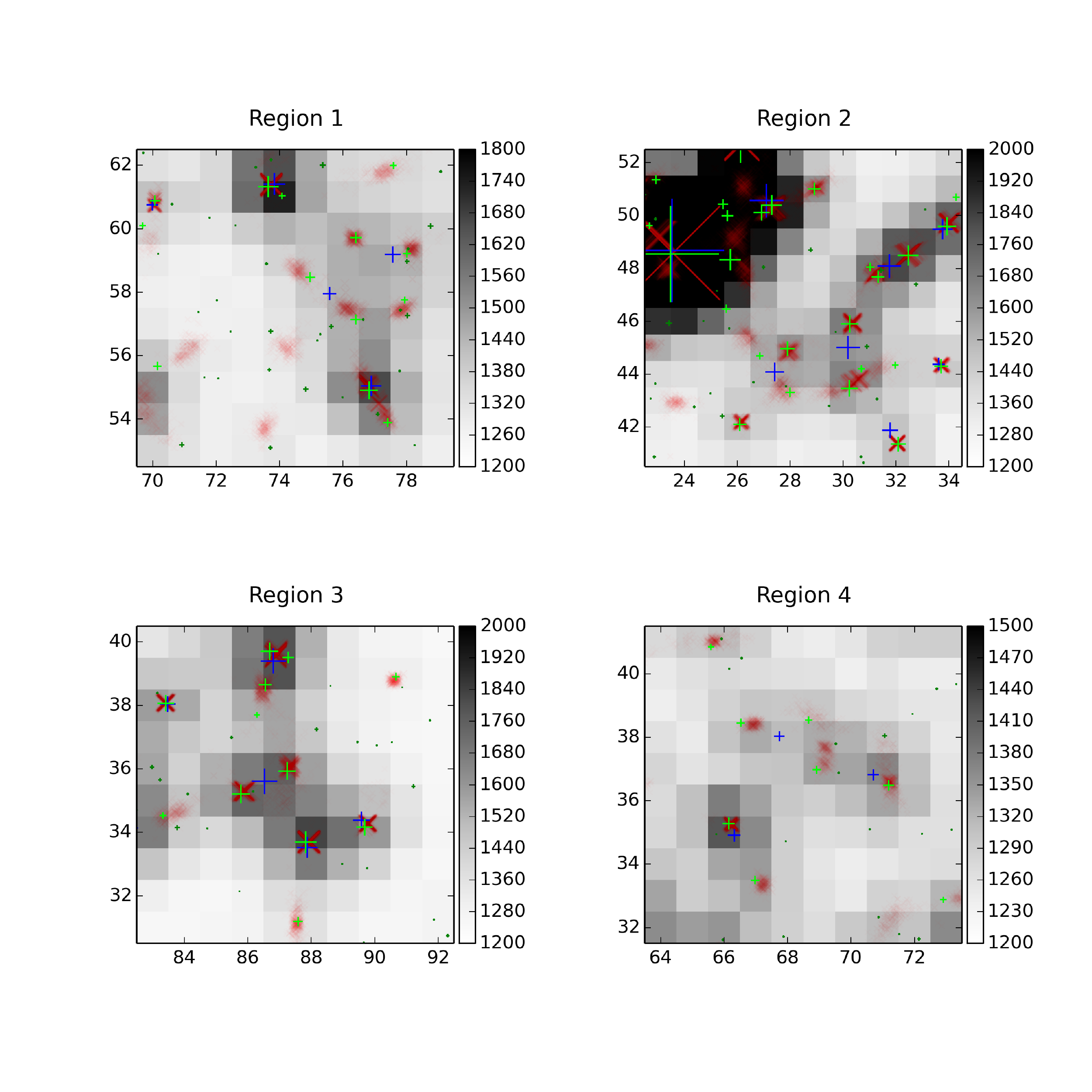}
\caption{\added{Depiction of the catalog ensemble in four 10x10 pixel cutouts of the SDSS image. Lime green crosses are HST sources brighter than 22\textsuperscript{nd} magnitude in F606W and dark green crosses are HST sources dimmer than this magnitude, both with area proportional to F606W flux. Blue crosses are DAOPHOT sources and translucent red Xs are sources from the catalog ensemble, stacked. The area of the DAOPHOT catalog and catalog ensemble source symbols is proportional to SDSS r band flux.}}
    \label{fig:four_panel}
\end{figure*}

\section{Recovering a Condensed Catalog}
\label{sec:condensed}
The burden of retaining and processing a large number of catalog samples may be a serious obstacle to using probabilistic cataloging. There is little advantage to a catalog ensemble if sources are unambiguously detected, but even in cases with severe crowding, a large fraction of the sources may be unambiguous. In this case, it may be useful to reduce the catalog ensemble to a single catalog, which we term a condensed catalog. This catalog has the convenience of a traditional catalog of one entry per object with uncertainties on each parameter, but retains the advantage of a fully marginalized posterior. That is, the flux and position uncertainties account for the effects of close neighbors. This middle road between traditional methods and a full catalog ensemble may be advantageous for many applications, especially if the user is most interested in properties of the brightest sources properly marginalized over the faint sources. In this section we propose a way to recover a condensed catalog from a catalog ensemble, and assess the condensed catalog's performance. 

\subsection{Labeling Sources}
\label{sec:labeling}
The main complication in reducing a catalog ensemble to a condensed catalog is that the probabilistic catalog's prior and likelihood are invariant under relabeling of the sources. While the MCMC sampling works by perturbing the properties of an ordered list of sources, the ordering of sources in this list cannot be taken as a consistent labeling of sources across samples in the catalog ensemble. It is useful to assign labels across samples to sources that are confidently detected in the catalog ensemble (i.e. they appear with well-constrained properties in a significant fraction of the samples). This labeling allows the position and flux distribution of these sources to be reported, just as in a traditional catalog. With a labeling procedure that at least works for unambiguous sources, we can evaluate the catalog ensemble's false discovery rate for the sources it infers confidently. Labeling becomes more difficult for faint and crowded sources; however, the catalog ensemble makes less confident inferences about these sources, and its false discovery rate for low-confidence inferences is less problematic. We present a simple labeling procedure that recovers the sources that the probabilistic catalog confidently infers.

Our labeling procedure starts by constructing a seed catalog that denotes likely positions for \added{sources} identified by the catalog ensemble. First, the sources from all of the samples in the catalog ensemble are stacked. A given source is considered a neighbor to the closest source from each of the other samples in the ensemble within 0.75 pixels, if such a source exists for that other sample. Then, the source with the most neighbors is denoted a seed (if there is a tie, a source is chosen arbitrarily). The new seed and its neighbors are removed from the seeding algorithm, preventing them from becoming future seeds and decrementing the neighbor count of their neighbors. The procedure of identifying seeds and removing them and their neighbors is repeated until all sources have been exhausted.

Then, we consider the brightest source in the seed catalog and identify the brightest matched source from each of the probabilistic catalog samples within a search radius (0.75 pixels), if there is such a source. Each match is labeled as corresponding to the brightest seed catalog source. These matched sources are eliminated from the matching algorithm, as they now have labels. This matching is then repeated for the second brightest source in the seed catalog, then the third brightest, etc. until all sources in the seed catalog have been considered. Using these labels, a condensed catalog can be produced containing a mean position and flux for each source, with error bars, as well as a prevalence  (the fraction of samples containing this source).

\subsection{Comparing to Catalog Ensemble}

\begin{figure}
\plotone{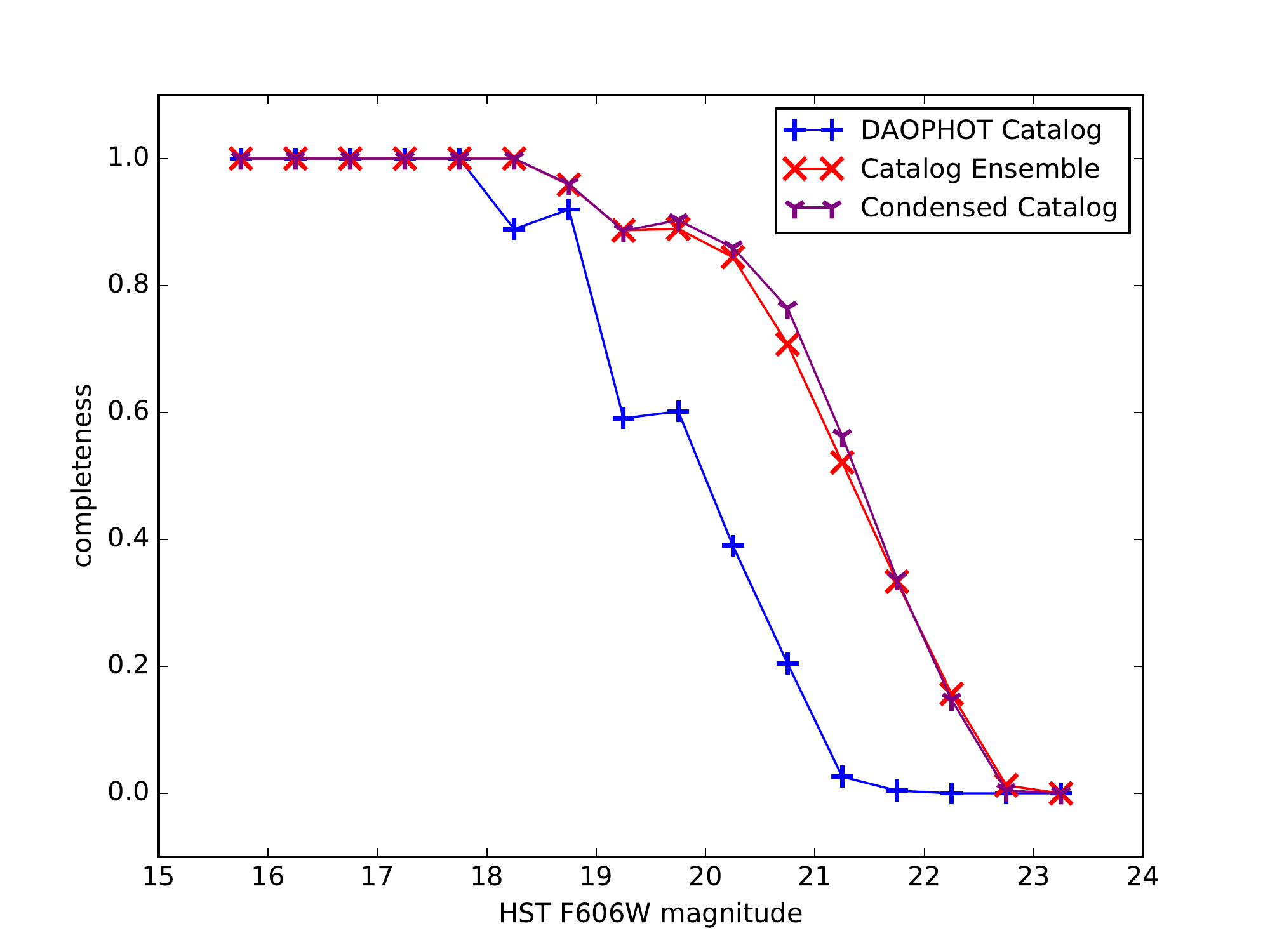}
\caption{Completeness of the condensed catalog compared to the DAOPHOT catalog and catalog ensemble.}
    \label{fig:completeness_condensed}
\end{figure}

The condensed catalog naturally marginalizes each labeled source's properties over the properties of nearby sources. However, in reporting independent distributions for each source, the condensed catalog contains less information than does the catalog ensemble. The condensed catalog does not capture source-source covariance like the catalog ensemble does, but it does reflect the impact that source-source covariance has on the uncertainty in each source's properties. We investigate the impact of this information loss on the completeness of the condensed catalog against the HST catalog. We generalize the notion of completeness to condensed catalogs. We do so by counting a condensed catalog source with a prevalence $p$ that matches with an HST source not as one match but as a fraction $p$ of a match. If, for all HST sources, all of the catalog ensemble sources that matched that HST source carry the same label and no other sources carry this label, then this completeness will match the catalog ensemble completeness. In Figure \ref{fig:completeness_condensed}, we present a comparison of the completeness of the probabilistic catalog, condensed catalog, and the DAOPHOT catalog. The condensed catalog performs almost identically to the probabilistic catalog, demonstrating that the labeling is stable for at least the sources that correspond to HST sources.

\begin{figure}
\plotone{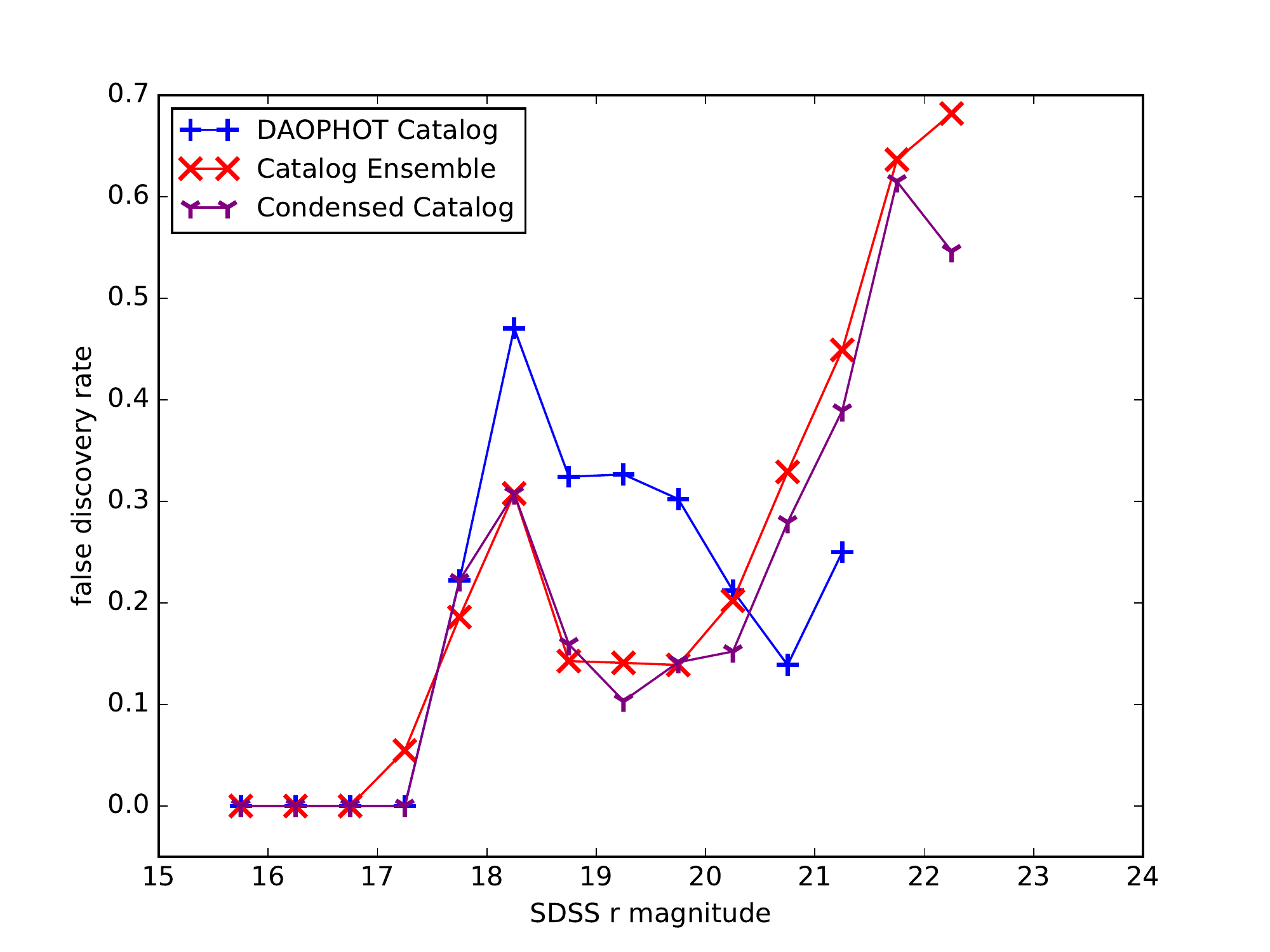}
\caption{False discovery rate of the condensed catalog compared to the DAOPHOT catalog and catalog ensemble.}
    \label{fig:falsediscovery_condensed}
\end{figure}

We also characterize the condensed catalog's false discovery rate as compared to the catalog ensemble. To generalize the false discovery rate (Equation \ref{eqn:fdr}) to condensed catalogs, we count a condensed catalog source with a prevalence $p$ as a fraction $p$ of a true positive if it matches an HST source, or as $p$ of a false positive if it does not.  If, for all HST sources, all of the catalog ensemble sources that matched that HST source carry the same label and no other sources carry this label, then this false discovery rate will match the catalog ensemble false discovery rate. The false discovery rate of the condensed catalog is compared to that of the catalog ensemble and DAOPHOT catalog in Figure \ref{fig:falsediscovery_condensed}. The condensed catalog's false discovery rate is comparable to that of the catalog ensemble's, showing that labeling does not greatly affect the false discovery rate.

\subsection{Completeness\textemdash False Discovery Tradeoff}

\begin{figure}
\plotone{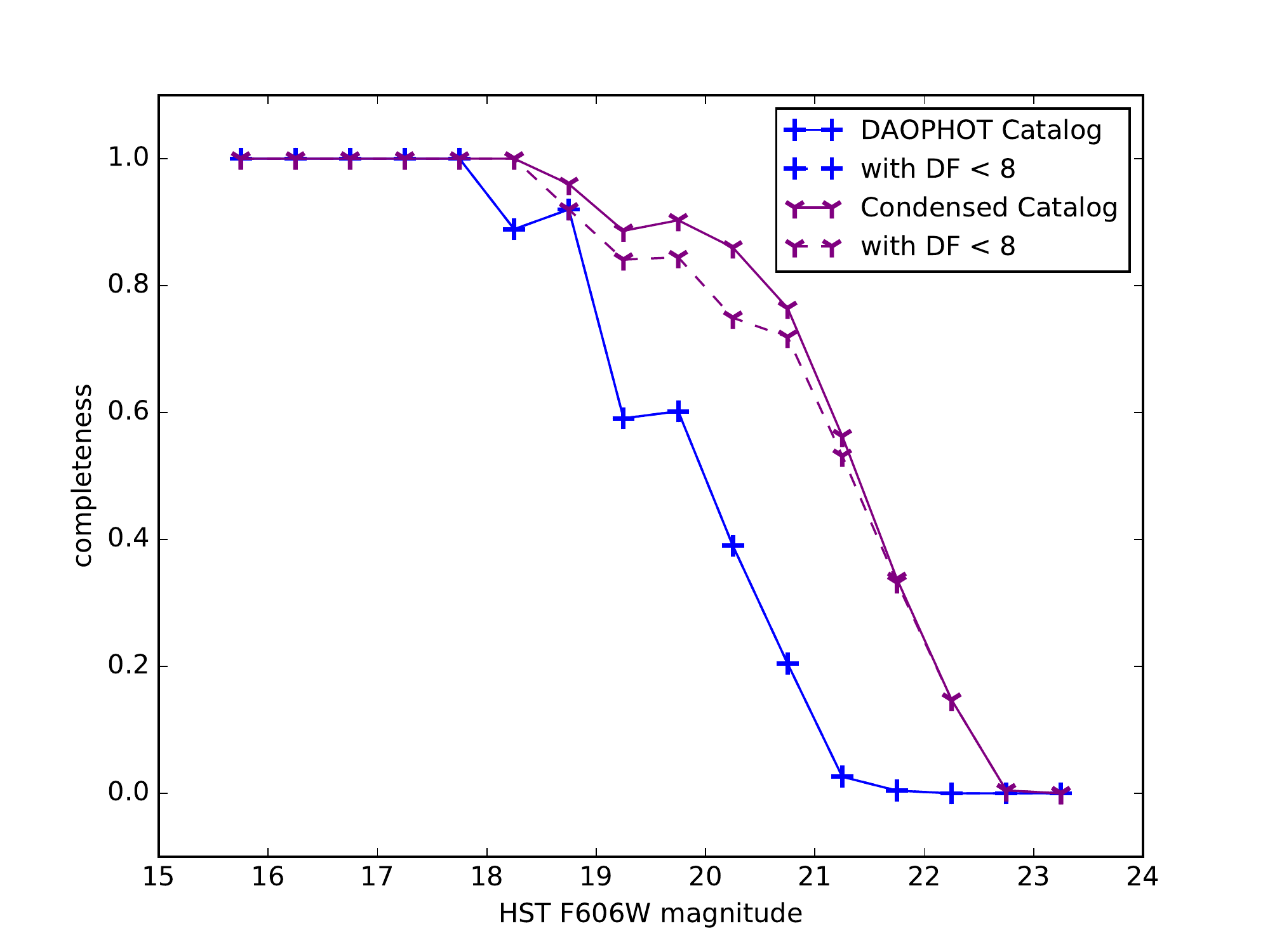}
\caption{Completeness of the DAOPHOT and condensed catalogs, before and after a loose \replaced{sigfac}{degradation factor} cut.}
    \label{fig:completeness_sigfac}
\end{figure}

\begin{figure}
\plotone{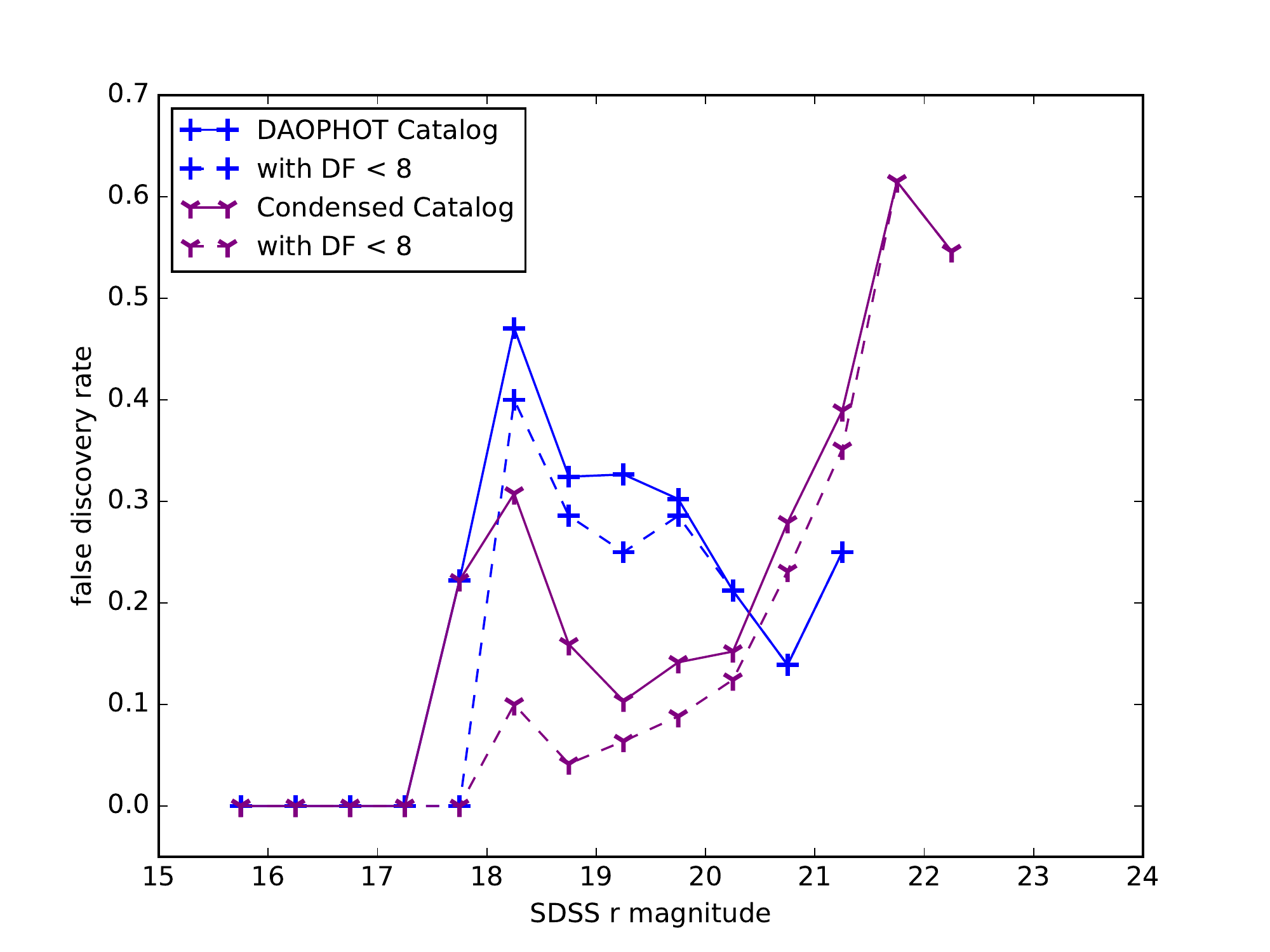}
\caption{False discovery rate of the DAOPHOT and condensed catalogs, before and after a loose \replaced{sigfac}{degradation factor} cut.}
    \label{fig:falsediscovery_sigfac}
\end{figure}

To help identify spurious sources and sources heavily contaminated by their neighbors, we compare the errors reported in the condensed catalog to the errors that would be expected for a given source based on its flux in a sparse field (see Appendix \ref{sec:sigfac} for the derivation). Sources that are contaminated by their neighbors will have larger errors than they would in a sparse field. Additionally, sources that are not well fit by our model because they are unresolved blends or artifacts of PSF mismatch will also have larger errors than expected. Cutting sources with a high ratio of reported to expected error (denoted \replaced{``sigfac'' here as an abbreviation for ``sigma factor''}{here as the ``degradation factor'' and abbreviated as DF}) may then remove many spurious sources while losing few legitimate sources. \explain{We decided that ``degradation factor'' was a better name than ``sigma factor'' for the ratio of actual error to expected error.} With a very loose cut at a flux \replaced{sigfac}{degradation factor} of 8, the false positive rate falls dramatically between 17\textsuperscript{th} and 20\textsuperscript{th} magnitude while only modestly decreasing completeness. \replaced{Sigfac in flux}{The flux degradation factor} can also be calculated for the DAOPHOT catalog based on the reported magnitude errors. Note that the DAOPHOT flux errors are not marginalized over neighboring sources, so DAOPHOT sources generally have lower \replaced{sigfac}{degradation factors} and are thus less affected by \replaced{sigfac}{degradation factor} cuts. Making the same cut at \replaced{sigfac}{degradation factor} 8 decreases the false positive rate of the DAOPHOT catalog without visibly impacting the completeness. Still, the DAOPHOT catalog's false discovery rate is higher than that of the condensed catalog with the same \replaced{sigfac}{degradation factor} cut. We plot the completeness for the condensed catalog and DAOPHOT catalog, with and without this \replaced{sigfac}{degradation factor} cut, in Figure \ref{fig:completeness_sigfac}, and we plot the false discovery rates in Figure \ref{fig:falsediscovery_sigfac}.

\begin{figure}
\plotone{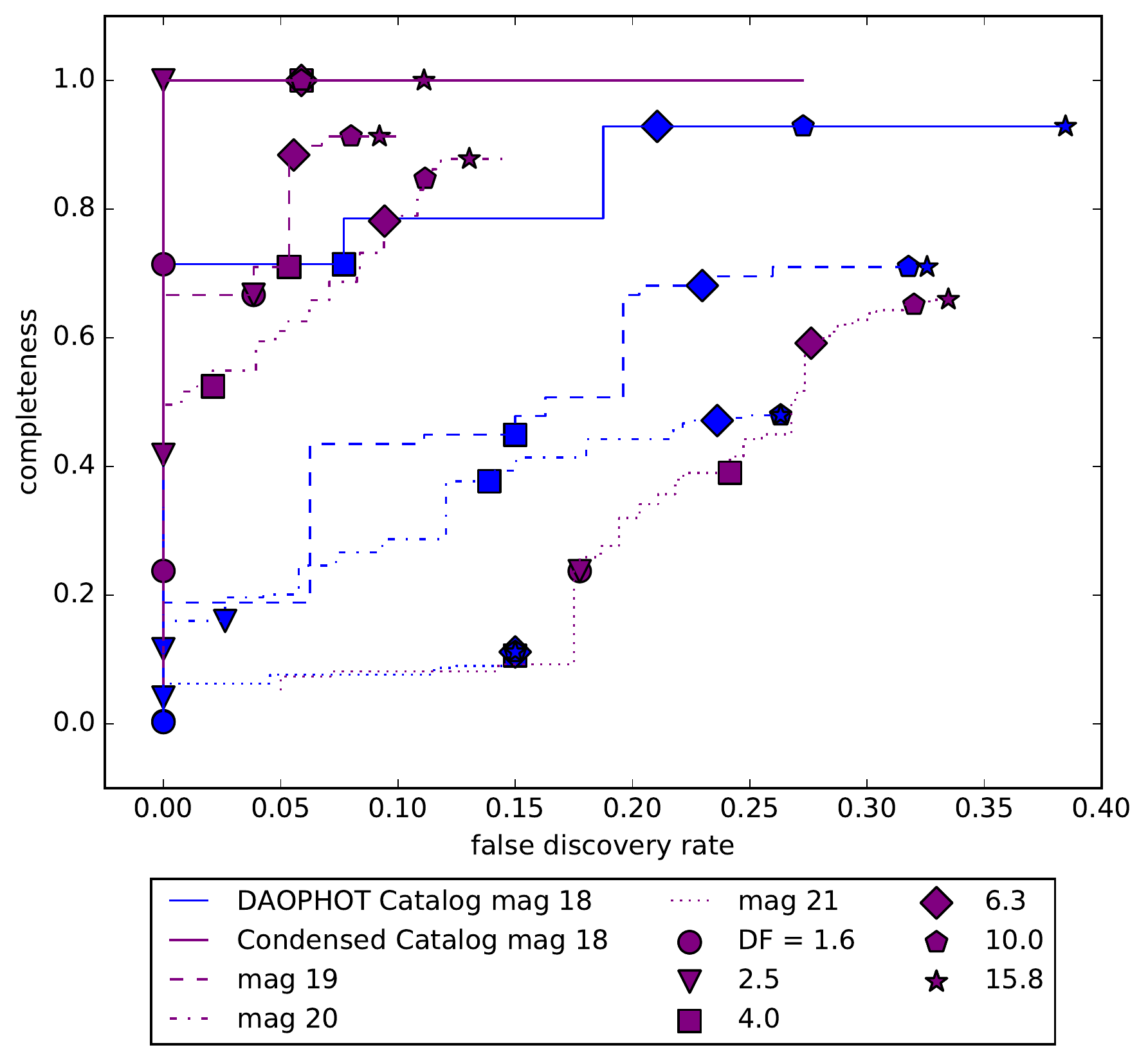}
\caption{Completeness-false discovery rate tradeoff of the DAOPHOT (blue lines) and condensed (purple lines) catalogs using cuts on \replaced{sigfac}{degradation factor}, evaluated at different magnitudes. Allowing the \replaced{sigfac}{degradation factor} cut to vary traces a curve in the completeness-false discovery plane. Several \replaced{sigfac}{degradation factor} cut values on each curve are denoted by symbols. Better performance lies closer to the upper left corner of the plot.}
    \label{fig:roc}
\end{figure}

Cuts on parameters like \replaced{sigfac}{the degradation factor} can be thought of as tuning parameters for a catalog. More permissive cuts allow more real sources to be identified, but at the risk of allowing more false positives. Conversely, more restrictive cuts decrease the number of false positives, but at the cost of losing completeness. Inspired by receiver operating characteristic curves in the binary classification literature, we plot the performance of the DAOPHOT and condensed catalogs in the space of completeness versus false discovery rate. Varying the \replaced{sigfac}{degradation factor} cut for a given catalog, and thus making a different trade off between high completeness and low false discovery rate, traces a curve in this space. A perfect cataloger would lie at the top left \textemdash full completeness with no false positives. A cataloger whose curve lies completely above and to the left of another cataloger's curve is superior in both metrics. We report the completeness versus false discovery rate curves at 18\textsuperscript{th}, 19\textsuperscript{th}, 20\textsuperscript{th}, and 21\textsuperscript{st} magnitudes in Figure \ref{fig:roc}. The condensed catalog clearly outperforms the DAOPHOT catalog, except at 21\textsuperscript{st} magnitude if a false discovery rate of less than \replaced{7\%}{15\%} is demanded. However, the false discovery rate is not well-measured for these restrictive cuts that do not admit many source candidates: the DAOPHOT catalog false discovery rate is \replaced{1/22}{6/40} compared to the condensed catalog's \replaced{1/15}{5/35}. Other parameters could be used in addition to attempt to further improve the condensed catalog's performance. We defer a full study of the completeness-false discovery tradeoff obtained with parameters other than \added{the} flux \replaced{sigfac}{degradation factor} for future work.

\added{
\subsection{Evaluating Convergence}
Evaluating convergence in catalog space is difficult, especially because of the labeling degeneracy. We do not expect formal convergence in the full catalog space with its $N!$ different labelings. At best, we desire convergence for high prevalence sources, ignoring the labeling degeneracy. We use the condensed catalog to evaluate convergence, as the labeling procedure finds sources in different catalog samples that correspond to each other. Our labeling procedure only works well for confidently inferred sources, so the convergence diagnostic is most meaningful for these sources.

We run the probabilistic cataloger four times and calculate a Potential Scale Reduction Factor (PSRF) \citep{1992StaSc...7..457G} using the four condensed catalogs. The condensed catalogs are first filtered to leave only sources with prevalence above 95\%. Then, we start associating the catalogs by considering the brightest source in the first condensed catalog and find the brightest match within 0.2 pixels of it in each of the other condensed catalogs. These sources are removed from the association algorithm, and this matching is then repeated for the second brightest source in the first condensed catalog, then the third brightest in the first condensed catalog, etc. until all sources in the first condensed catalog have been associated. Each of $N_c=4$ condensed catalogs reports a mean $\mu_c$ and variance $\sigma_c^2$ for each source parameter, so the PSRF can be calculated for each associated source's parameters:
\begin{equation}
\label{eqn:PSRF}
PSRF = \sqrt{1+\frac{B}{W}-\frac{1}{N_s}}
\end{equation}
where $N_s$ is the number of samples used to make each condensed catalog, $W$ is the within-chain variance, ie. the mean of the parameter variances:
\begin{equation}
\label{eqn:withinchain}
W = \frac{1}{N_c} \sum_{c=1}^{N_c} \sigma_c^2,
\end{equation}
and B is the between-chain variance, ie. the variance of the chain means:
\begin{equation}
\label{eqn:betweenchain}
B = \frac{1}{N_c - 1} \sum_{c=1}^{N_c} \left(\mu_c - \frac{\sum_{c=1}^{N_c} \mu_c}{N_c}\right)^2.
\end{equation}
In some very crowded regions, the condensed catalogs disagree on how many sources are present. In these cases, the labeling procedure has not worked well because the sources are not significant or isolated enough, and so the calculated PSRF is not useful. To try to avoid these cases, we deem any condensed catalog source without another source within 1.5 pixels (twice the labeling match radius) to be `isolated' and thus have a trustworthy PSRF. This criterion is very conservative and excludes some sources that are well-labeled and thus have meaningful PSRFs. We plot a histogram of the PSRF of the fluxes of isolated and non-isolated sources in Figure \ref{fig:PSRF}. Most isolated sources have PSRFs close to 1, suggesting that the cataloger has converged on most of these sources. The highest PSRF for an isolated source is 1.3, suggesting that the sampler has not completely converged on some sources.

\begin{figure}
\plotone{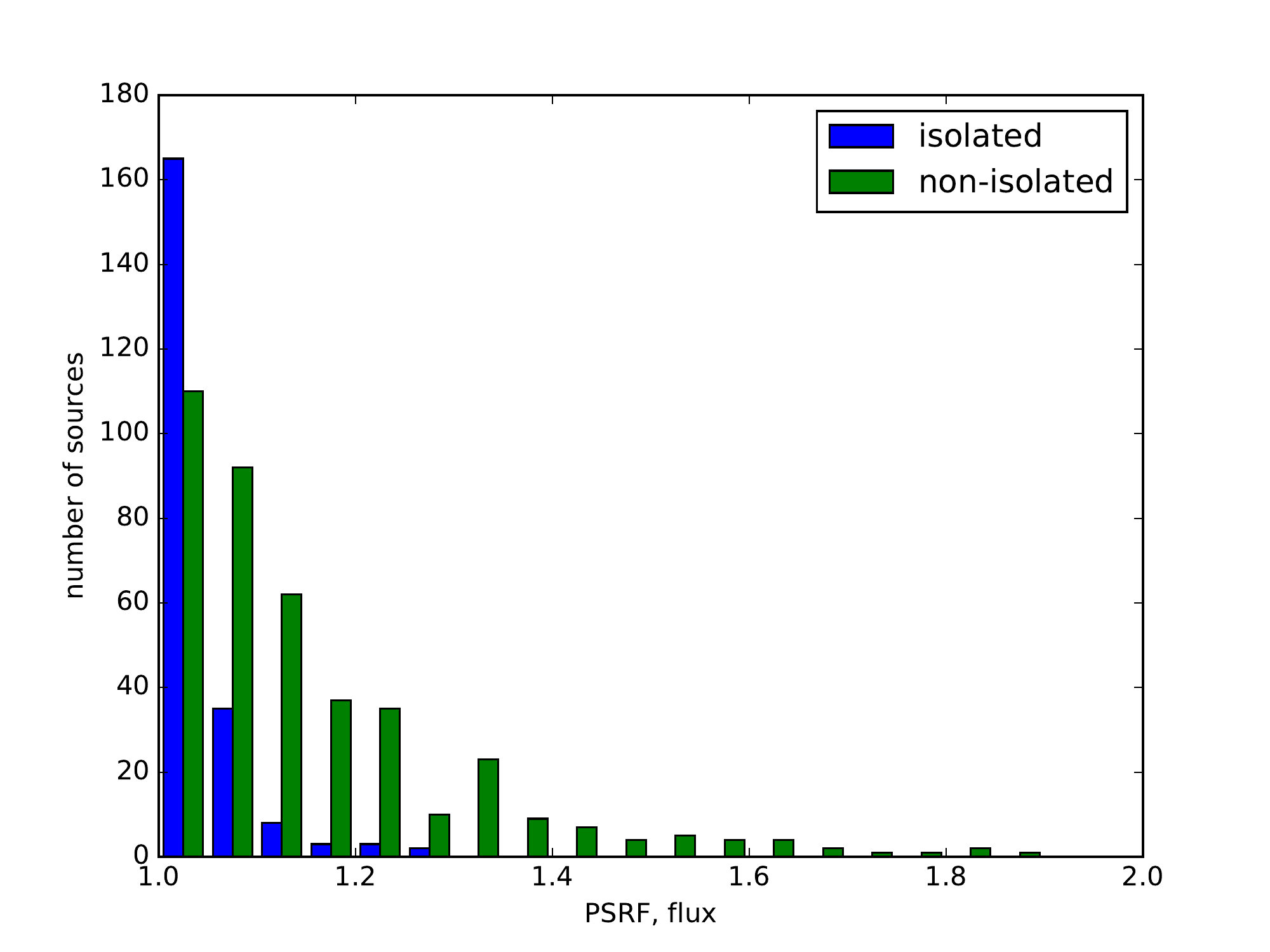}
\caption{\added{Potential Scale Reduction Factor for isolated and non-isolated sources.}}
    \label{fig:PSRF}
\end{figure}

}
\section{Discussion and Conclusion}
\label{sec:discussion}
We implement a probabilistic cataloger that produces an ensemble of point source catalogs of a crowded SDSS field in M2. We demonstrate that this catalog ensemble better recovers sources than the DAOPHOT catalog does by comparing to the sources identified by the HST catalog in the same field. The catalog ensemble is more than 1 magnitude deeper in completeness and has a lower false discovery rate above 20\textsuperscript{th} magnitude. Many closely spaced sources that are blended in the DAOPHOT catalog are deblended in the catalog ensemble, and the residuals of the catalog ensemble model image are smaller and less spatially structured than those of the DAOPHOT model image. DAOPHOT seeds candidate source locations using a peak finder on a PSF-convolved image, making it likely to blend closely spaced sources. By instead proposing new sources' locations and evaluating their likelihood, probabilistic cataloging is better able to separate closely spaced sources. In a crowded field, better deblending is a great advantage, explaining the catalog ensemble's superior performance in completeness, false discovery rate, and residuals.

Distilling the obtained catalog ensemble to a condensed catalog does not lose much of the information contained in the catalog ensemble about confidently inferred sources. The condensed catalog has comparable completeness and false discovery rate to the catalog ensemble. We define a metric of quality: \replaced{sigfac}{the degradation factor}, the ratio of the flux error to the expected flux error for a source if it were isolated. A loose \replaced{sigfac}{degradation factor} cut drastically reduces the false positive rate while not impacting completeness by removing sources with abnormally high error, including some artifacts of PSF mismatch and incorrectly blended sources. While the condensed catalog appears similar to a traditional catalog, the errors for a source in this condensed catalog have been marginalized over the possible properties of its neighbors and nuisance parameters like the sky level.

The labeling algorithm for the condensed catalog detailed in Section \ref{sec:labeling} works well because most of the sources inferred were confidently inferred. In choosing our prior on the number of sources, we filtered out less significant sources that would have been included with less confidence. Labeling probabilistic catalogs including these less confident sources will be more difficult.

\added{No single catalog can serve all purposes. Our aim is to produce a condensed catalog that goes deeper in crowded fields and has more principled errors, and a catalog ensemble that may be useful for some purposes. Any user wishing to impose substantially different priors will need to run the cataloger again. To that end, we are currently working on computational efficiency so that the sophisticated user might be presented with code to create their own catalogs.

In this work, we make some strong assumptions in order to simplify our probabilistic cataloging implementation. Firstly, we assume that the PSF is constant throughout the image and is already known. The image used is small enough that it is a good approximation to assume that the PSF is constant. For sparser fields, there are existing methods that could be used to extract a PSF estimate, so the PSF could be taken as known in these cases. However, extracting the PSF is more difficult in crowded fields. We use the PSF determined by the SDSS pipeline, which has been validated on sparser fields but not on fields as crowded as the one we consider. Looking at the residuals, the pipeline appears to have obtained a reasonable PSF estimate. The parameters of a PSF model could be included in the cataloging fit, as is done on mock data by \cite{2013AJ....146....7B}, capturing the degeneracies between the PSF and the catalog. We decided to fix the PSF for two reasons. First, the PSF from the SDSS pipeline matches the data well and captures features that cannot be described by a simple parameterization of the PSF. Second, it is desirable to have proposals that only require recalculating the model image for a small fraction of the sources in the image. Proposals which change the PSF require the model image for all sources to be recalculated, making these proposals much more computationally expensive. We investigate the effect of making the assumed PSF narrower and broader by 10\% in Appendix \ref{sec:psf_change} and find these perturbed PSFs significantly deteriorate the cataloger performance. Knowing the PSF is important for cataloging crowded fields, so it is necessary to either have robust PSF extraction methods or to fit the PSF parameters along with the catalog.

Secondly, we assume a uniform spatial prior for the sources. We use a uniform spatial prior because it simplifies the MCMC proposals and the cluster's spatial profile is not our primary interest. However, adding an analytic non-uniform prior would only add modestly to the computational complexity of the work. Given that we work with an image of a globular cluster, we could have instead used a hierarchical model with the central position and scale radius of the cluster as hyperparameters. The inferences for significant sources should not be greatly affected because the likelihood should dominate over any reasonable change to the prior. However, with a cluster hierarchical model, the catalog could be more sensitive to faint sources close to the center of the cluster (where the prior says there are more sources) and could be less likely to infer false positives in the outskirts (where the prior says there are fewer sources).

Thirdly, we assume that the flux distribution of sources is a power-law with a fixed index of 2, but in reality, the sources may (approximately) follow a different index or another distribution (eg. broken power-law). To reduce the amount of computational time spent on insignificant sources, we use Equation \ref{eqn:exponential_nprior} as our prior on $N$. However, this prior suppresses faint sources in a way we have not accounted for, making a hierarchical inference of the flux distribution untenable, so we use a flux distribution with no free parameters (see Appendix \ref{sec:power_law} for further discussion). The number of faint sources in the inferred catalog ensemble is affected by the assumed index: a steeper index gives more faint sources and a shallower index gives fewer, regardless of the actual underlying flux distribution. Bright, significant sources may be affected indirectly, as they may be oversplit more often (eg. due to imperfections in the PSF model) when the prior demands more faint sources than actually exist.

Finally, real data have some non-Gaussian noise like cosmic rays, diffraction spikes, and bleed trails. Because we simply aim to provide a proof of concept for probabilistic cataloging, we picked a region of interest that avoids such non-Gaussianties. Our code admits a pixel mask which can be used to mitigate these non-Gaussianities. In existing surveys, much work has been done on how to best handle these outliers.}

While our current implementation of probabilistic cataloging is only applicable to point sources, it can be generalized to extended sources like galaxies. In addition to position and flux, each source would have shape parameters (e.g. the parameters of an exponential or de Vaucouleurs profile) which would be used to calculate the model image against which the data are compared. Star-galaxy separation for galaxies that are not much bigger than the PSF may require special attention.

The main disadvantage of probabilistic cataloging is that it takes much more computing time than traditional cataloging. Given the amount of time needed to process the chosen $100\times100$ image, cataloging an optical sky survey would be impractical with the current implementation of probabilistic cataloging. However, the current implementation has not been optimized for computational speed, so substantial gains in speed can be expected as the implementation is further developed. Also, the proposals used in the MCMC are unoptimized and uninformed - studying how to make better proposals may significantly reduce the number of steps needed in the MCMC chain. Alternatively, there may exist some approximation of probabilistic cataloging that is much faster while still retaining its advantages in crowded fields.

As telescopes become increasingly sensitive and detect more sources, more of the sky will become crowded enough to pose problems for traditional cataloging algorithms. If probabilistic cataloging or some derivative of it can be sped up, near-future computing power may make it worth running on the most crowded fields of an optical survey, or perhaps the entire survey itself. In demonstrating that probabilistic cataloging has superior performance in crowded fields, we contend that it should be further developed in order to create a photometric pipeline that will be able to meet the challenges posed by the next generation of optical surveys.

\acknowledgments
SKNP is supported in part by a Natural Sciences and Engineering Research Council of Canada Postgraduate Scholarship.  BCGL was supported in part by the Harvard College Research Program.

We thank Brendon Brewer, Charles-Antoine Collins Fekete, Dan Foreman-Mackey, Gregory Green, Kiefer Hicks, Albert Lee, and Zachary Slepian for useful discussions during the course of the project.

The computations in this paper were run on the Odyssey cluster supported by the FAS Division of Science, Research Computing Group at Harvard University. This research has made use of NASA's Astrophysics Data System.

Funding for SDSS-III has been provided by the Alfred P. Sloan Foundation, the Participating Institutions, the National Science Foundation, and the U.S. Department of Energy Office of Science. The SDSS-III web site is \url{http://www.sdss3.org/}.

SDSS-III is managed by the Astrophysical Research Consortium for the Participating Institutions of the SDSS-III Collaboration including the University of Arizona, the Brazilian Participation Group, Brookhaven National Laboratory, University of Cambridge, Carnegie Mellon University, University of Florida, the French Participation Group, the German Participation Group, Harvard University, the Instituto de Astrofisica de Canarias, the Michigan State/Notre Dame/JINA Participation Group, Johns Hopkins University, Lawrence Berkeley National Laboratory, Max Planck Institute for Astrophysics, Max Planck Institute for Extraterrestrial Physics, New Mexico State University, New York University, Ohio State University, Pennsylvania State University, University of Portsmouth, Princeton University, the Spanish Participation Group, University of Tokyo, University of Utah, Vanderbilt University, University of Virginia, University of Washington, and Yale University.

\bibliographystyle{yahapj}
\bibliography{references.bib}

\appendix

\section{MCMC Proposals}
\begin{enumerate}
\item Choose a number of sources to perturb, and then perturb their positions and fluxes.
\item Perturb flux distribution parameters, changing all source fluxes to remain at the same quantile.
\item Perturb flux distribution parameters, keeping all source fluxes fixed.
\item Choose a number of sources to add, and then add sources drawn from the prior with the current flux distribution parameters.
\item Choose a number of sources to remove, and then remove sources chosen at random.
\end{enumerate}

Whenever a number of sources is chosen, a multi-scale distribution is used with a minimum of 1 and a maximum of order the maximum number of sources allowed (3,000 in this work). Whenever a parameter is perturbed, a multi-scale distribution is used with a smallest step of order $10^{-6}$ the prior width and a largest step of order the prior width. \added{These proposals are very similar to those used by \cite{2013AJ....146....7B}, except that we perturb source positions and fluxes in the same step while \cite{2013AJ....146....7B} perturbs positions and fluxes separately.}

\section{Sparse Field Validation}
\begin{figure}
\plotone{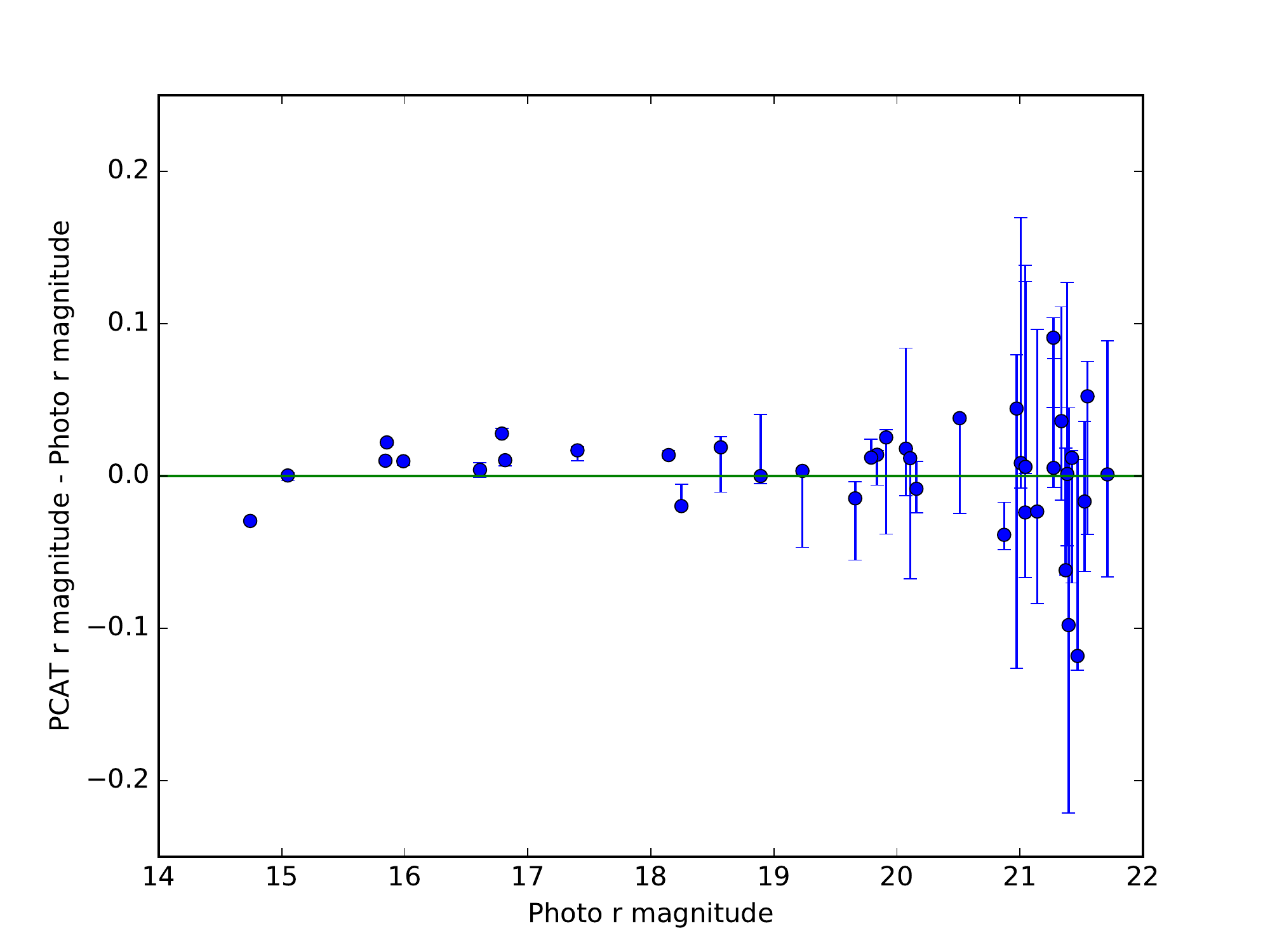}
\caption{Deviations of PCAT magnitudes with respect to Photo magnitudes in a sparse field, with 68\% credible intervals.}
    \label{fig:sparse_flux}
\end{figure}

\begin{figure}
\plotone{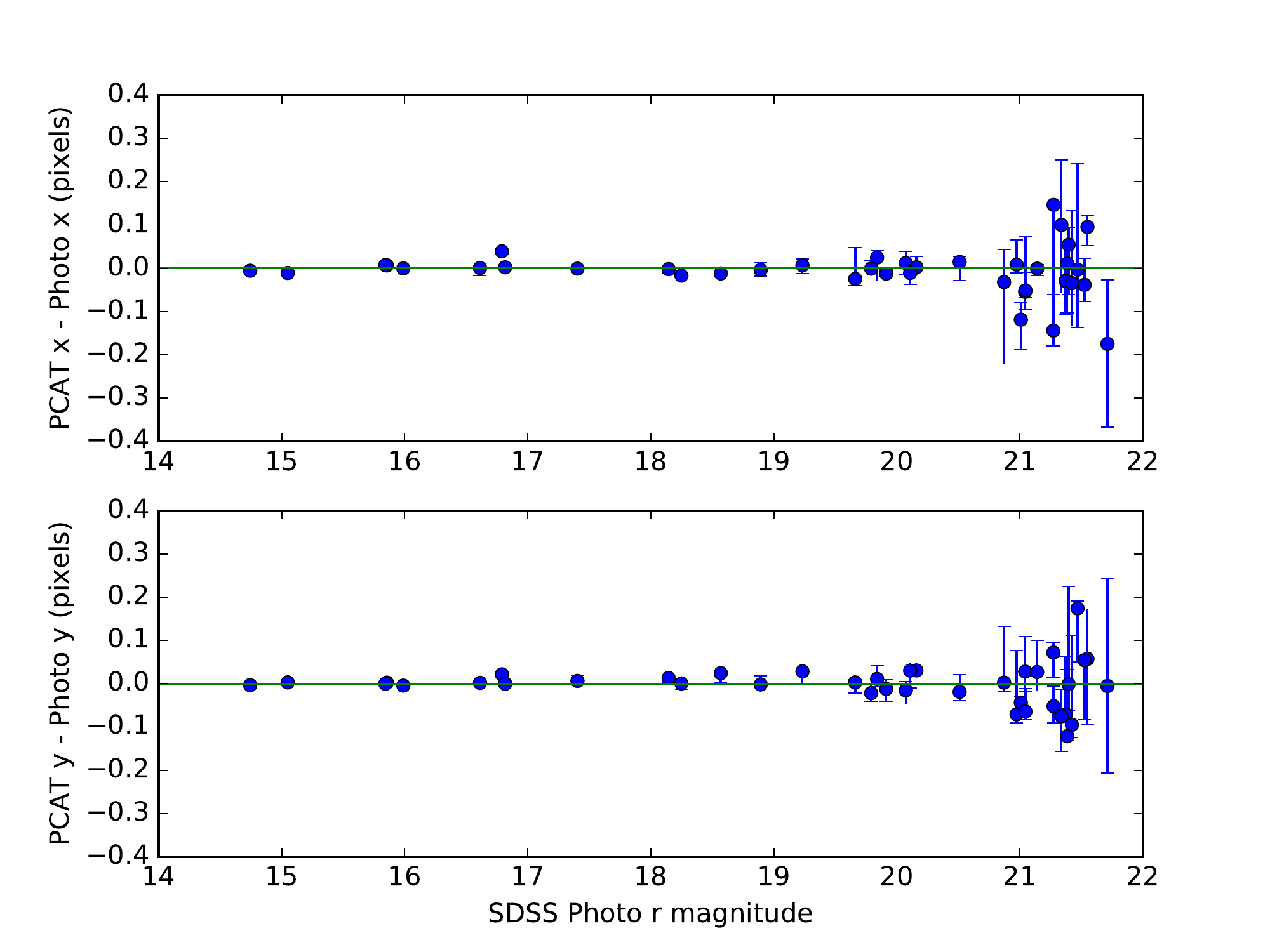}
\caption{Deviations of mean PCAT positions with respect to Photo positions in a sparse field.}
    \label{fig:sparse_pos}
\end{figure}

To validate our photometry, we also run our probabilistic cataloger on a sparse field where SDSS Photo performs well. A significant fraction of sources in this sparse field are galaxies, and thus extended. Our cataloger assumes all sources are point sources, so we only compare photometry for sources that Photo identifies as point sources. We find all of the point sources that Photo finds, and our photometry agrees with Photo to 0.015 mag at the bright end, growing to \replaced{0.09}{0.047} mag by 20\textsuperscript{th} magnitude (see Figure \ref{fig:sparse_flux}). The positions that we infer agree with the Photo positions to \replaced{0.06}{0.012} pixels at the bright end and \replaced{0.45}{0.058} pixels by 20\textsuperscript{th} magnitude (see Figure \ref{fig:sparse_pos}).

\section{Degradation Factor}
\label{sec:sigfac}
\explain{Renamed this section ``Degradation Factor''}
To simplify the derivation of the expected flux error, we will assume that the likelihood dominates over the prior and that the approximations made in Equation \ref{eqn:simpleloglike} hold. Assuming that the position of this source, the positions and fluxes of the other sources, and the nuisance parameters are well known, the maximum likelihood flux for a single source $f_{*}$ can be found by maximizing the log likelihood:

\begin{equation}
\label{eqn:maxlike}
\left.\frac{\partial \log \mathcal{L}}{\partial f}\right\vert_{f = f_{*}} = 0
\end{equation}

The flux variance about this maximum likelihood value is then:

\begin{equation}
\label{eqn:secondmoment}
\sigma_f^2 = \left<\left(f-f_{*}\right)^2\right> = \frac{\int_{-\infty}^{\infty} \exp\left(\log \mathcal{L}\right) (f-f_{*})^2 df}{\int_{-\infty}^{\infty} \exp\left(\log \mathcal{L}\right) df}
\end{equation}

Taylor expanding the log likelihood about $f_{*}$ to second order and using Equation \ref{eqn:maxlike} yields:

\begin{equation}
\label{eqn:fluxvariance}
\sigma^2_f \approx \frac{\int_{-\infty}^{\infty} \exp\left(\log \mathcal{L}\vert_{f=f_{*}} + \frac{1}{2} \left.\frac{\partial^2 \log \mathcal{L}}{\partial f^2} \right\vert_{f=f_{*}} (f-f_{*})^2 \right) (f-f_{*})^2 df}{\int_{-\infty}^{\infty} \exp\left(\log \mathcal{L}\vert_{f=f_{*}} + \frac{1}{2} \left.\frac{\partial^2 \log \mathcal{L}}{\partial f^2} \right\vert_{f=f_{*}} (f-f_{*})^2 \right) df} = \left(-\left.\frac{\partial^2 \log \mathcal{L}}{\partial f^2} \right\vert_{f=f_{*}}\right)^{-1}
\end{equation}

Assuming this source is well isolated, the model flux is $\lambda_{lm} \approx f \mathcal{P}(l-x, m-y) + I_{sky}$. Taking the second derivative of Equation \ref{eqn:simpleloglike} yields:

\begin{equation}
\label{eqn:secondderivativeflux}
\frac{\partial^2 \log \mathcal{L}}{\partial f^2} \approx - \sum_{l=1}^{W} \sum_{m=1}^{H} \frac{\mathcal{P}^2(l - x, m - y)}{ \lambda_{lm}} \left(1 + \frac{k_{lm} - \lambda_{lm}}{\lambda_{lm}}\right)^2
\end{equation}

Assuming the fractional deviations $(k_{lm} - \lambda_{lm})/\lambda_{lm}$ to be small gives:

\begin{equation}
\label{eqn:secondderivativefluxsimple}
\frac{\partial^2 \log \mathcal{L}}{\partial f^2} \approx - \sum_{l=1}^{W} \sum_{m=1}^{H} \frac{\mathcal{P}^2(l - x, m - y)}{f \mathcal{P}(l-x, m-y) + I_{sky}}
\end{equation}

Substituting this expression into Equation \ref{eqn:fluxvariance} gives

\begin{equation}
\label{eqn:fluxerror}
\sigma^2_f \approx \left(\sum_{l=1}^{W} \sum_{m=1}^{H} \frac{\mathcal{P}^2(l - x, m - y)}{f_{*} \mathcal{P}(l-x, m-y^{*}) + I_{sky}} \right)^{-1}
\end{equation}

\begin{figure}
\plotone{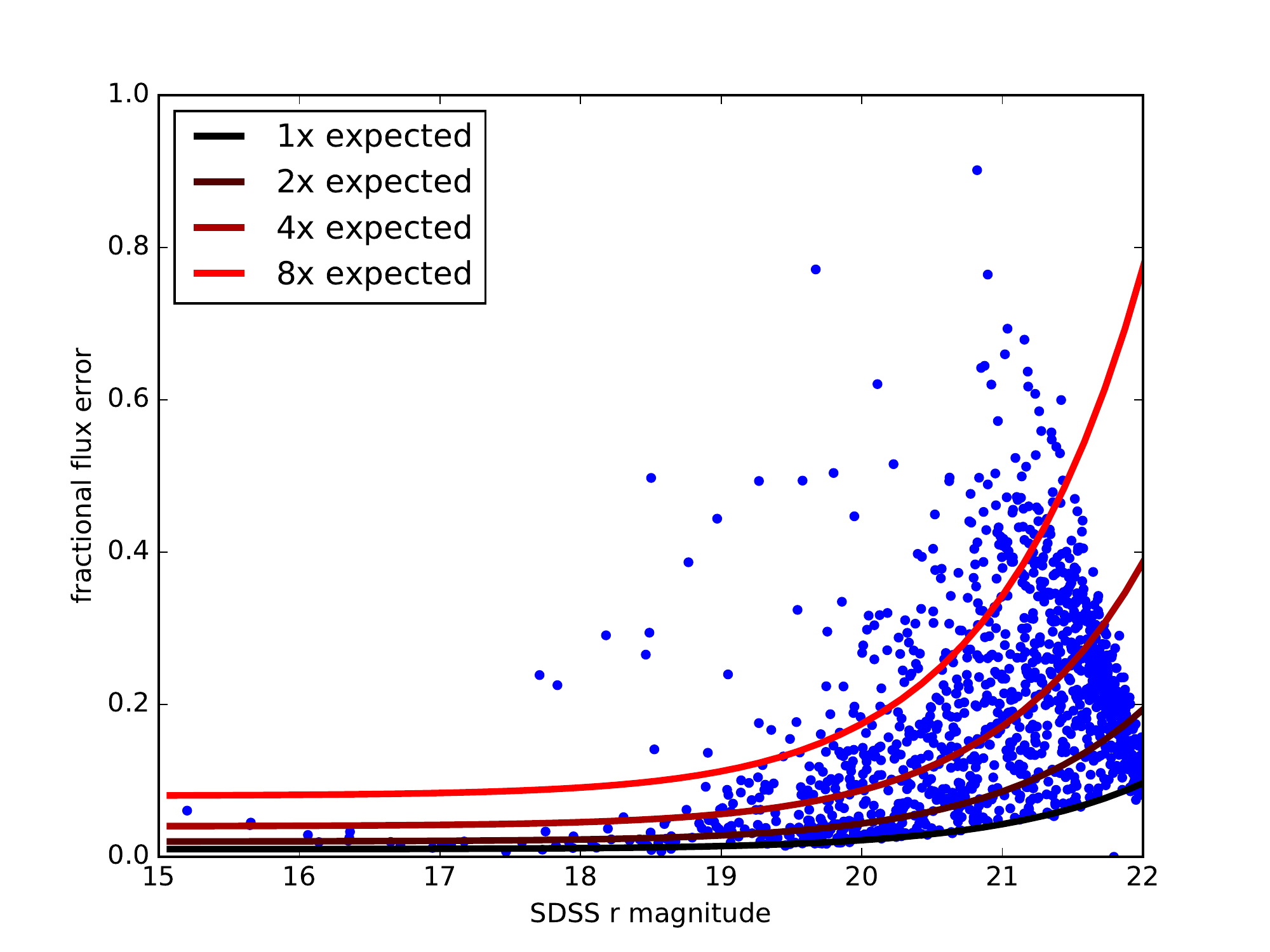}
\caption{Flux errors reported in the condensed catalog compared to errors expected in the sparse field limit.}
    \label{fig:sigfac_flux}
\end{figure}

The reported errors in the condensed catalog are plotted in Figure \ref{fig:sigfac_flux}. Adding a noise floor of 1\% to account for effects like PSF mismatch which this derivation ignores, the expected flux error gives an excellent lower bound to the reported flux errors. Some sources near $f_{min}$ appear to have smaller flux errors because their flux distributions are cut off at the faint end by $f_{min}$. Because of how crowded the field is, many sources have flux errors many times larger than expected in the sparse limit.

By a similar procedure, the position uncertainty can be found by first taking the second derivative of Equation \ref{eqn:simpleloglike} with respect to $x$ or $y$:

\begin{equation}
\label{eqn:secondderivativeposition}
\frac{\partial^2 \log \mathcal{L}}{\partial x^2} \approx - f_{*}^{2} \sum_{l=1}^{W} \sum_{m=1}^{H} \frac{1}{\lambda_{lm}} \left(\left[1 + \left(\frac{k_{lm}-\lambda_{lm}}{\lambda_{lm}}\right)^2\right] \left[\frac{\partial \mathcal{P}}{\partial x}(l-x, m-y_{*})\right]^2 - \frac{\lambda_{lm}}{f_{*}}\frac{k_{lm} - \lambda_{lm}}{\lambda_{lm}}\left[1+\frac{1}{2} \frac{k_{lm}-\lambda_{lm}}{\lambda_{lm}}\right] \frac{\partial^2 \mathcal{P}}{\partial x^2}(l-x, m-y_{*}) \right)
\end{equation}

The second term is suppressed by the pixel model flux $\lambda_{lm}$ divided by the entire flux of the star $f_{*}$. Dropping this term and assuming the fractional deviations $(k_{lm} - \lambda_{lm})/\lambda_{lm}$ to be small gives:

\begin{equation}
\label{eqn:positionerror}
\sigma^2_x \approx \frac{1}{f_{*}^{2}} \left(\sum_{l=1}^{W} \sum_{m=1}^{H} \frac{1}{f_{*} \mathcal{P}(l-x_{*}, m-y_{*}) + I_{sky}} \left[\frac{\partial \mathcal{P}}{\partial x}(l-x_{*}, m-y_{*})\right]^2 \right)^{-1}
\end{equation}

\begin{figure}
\plotone{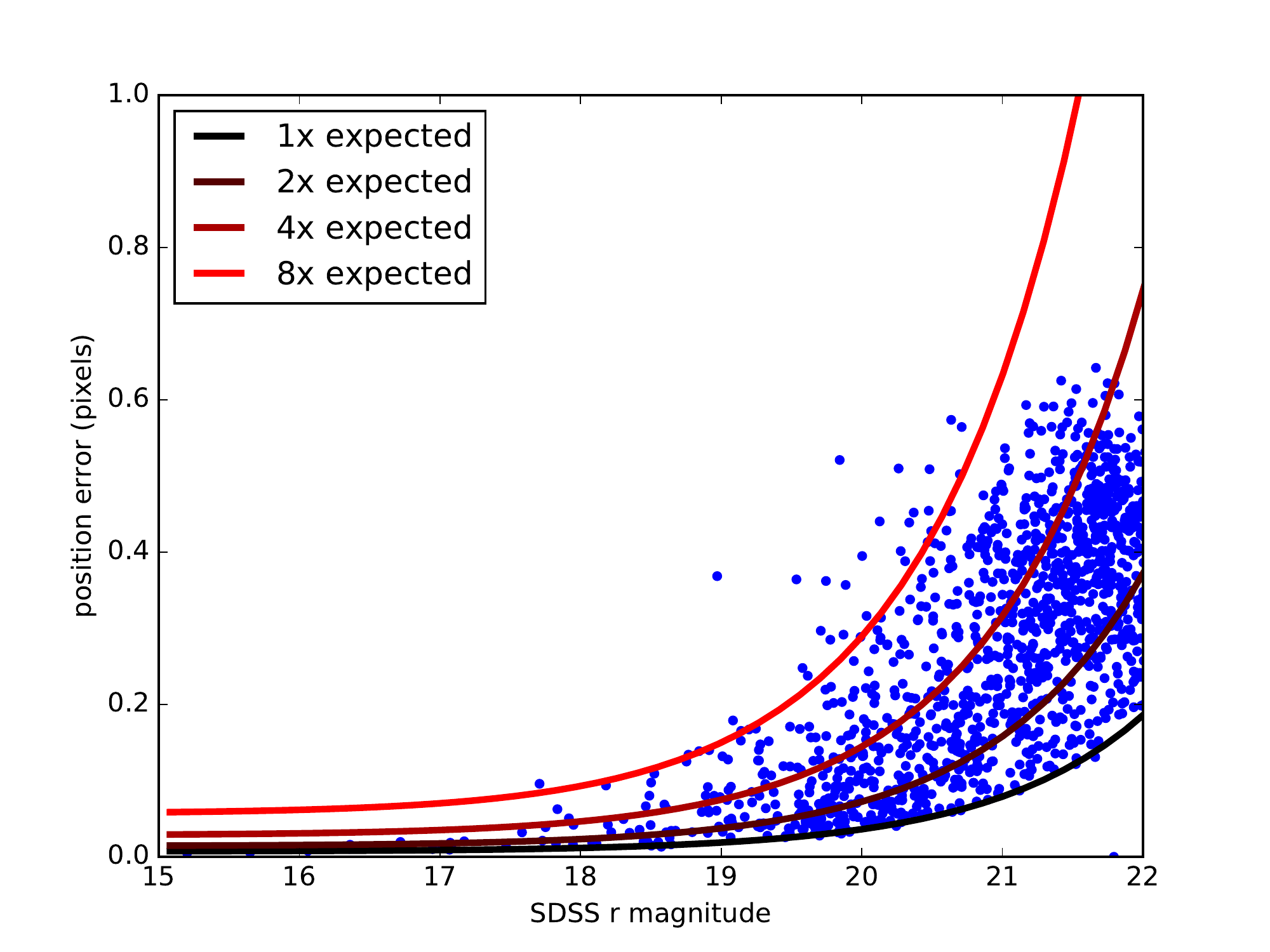}
\caption{Position errors reported in the condensed catalog compared to errors expected in the sparse field limit.}
    \label{fig:sigfac_pos}
\end{figure}

Adding a noise floor of 0.005 pixels in both $x$ and $y$, the expected position error $\sqrt{\sigma_x^2 + \sigma_y^2}$ provides an excellent lower bound to the reported position errors as seen in Figure \ref{fig:sigfac_pos}.

\begin{figure}
\plotone{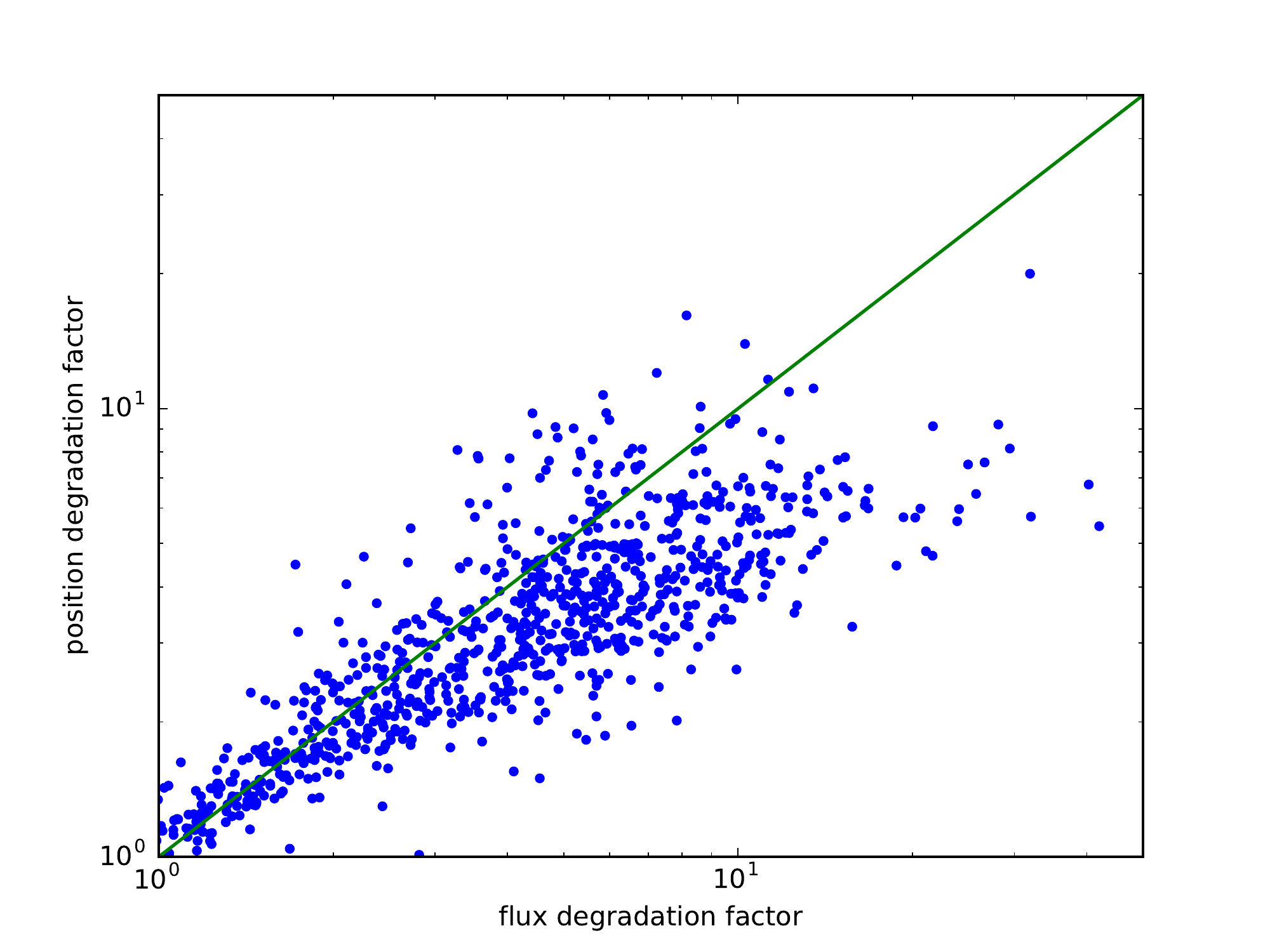}
\caption{Correlation between position \replaced{sigma}{degradation} factors and flux \replaced{sigma}{degradation} factors reported in the condensed catalog.}
    \label{fig:sigfac_corr}
\end{figure}

The \replaced{sigfacs}{degradation factors} (ratio of reported error to expected error) in flux and position are correlated but not redundant, as seen in Figure \ref{fig:sigfac_corr}.

\added{
\section{Sensitivity to Flux Distribution Power-Law Index}
\label{sec:power_law}
We fix the index of the flux distribution power-law because the exponential prior on $N$ (Equation \ref{eqn:exponential_nprior}) suppresses the number of dim sources. The dimmer a source is, the smaller is the gain in likelihood from including it and the less likely it is to be included under an exponential prior on $N$. This suppression of dim sources means the sub-threshold population of sources cannot be inferred, but the likelihood evaluation is faster because fewer sources are being included. The focus of this work is the deblending of significant sources, so the suppression of dim sources is an acceptable side effect. Allowing the power-law index to vary would merely fit this dim end turn off, not the power-law of the underlying population. If this suppression of dim sources could be predicted, it could be accounted for in the prior on individual source fluxes, and the power-law index of the actual population could then be inferred. We leave a calculation of this suppression for future work.

The prior on $N$ and even the choice of power-law index should not greatly affect significant sources, which are the focus of this work. The large gains in likelihood from including significant sources ensures that they will be included, regardless of the choice of prior. As a cross-check, we create catalog ensembles assuming power-law indices of 1.5 and 2.5 to compare to the power-law index of 2.0 used in this work. The flux distributions of the catalog ensemble sources using all three indices are plotted in Figure \ref{fig:alpha_dist}. All the catalog ensembles agree at the bright end, but the higher index catalog ensembles have more sources at the dim end. If we use the condensed catalogs derived from these catalog ensembles and only consider sources with over 95\% prevalence, the condensed catalogs are in good agreement, as shown in Figure \ref{fig:alpha_prevalence}. The completenesses and false discovery rates of these catalog ensembles are compared in Figures \ref{fig:completeness_alpha} and \ref{fig:fdr_alpha}, respectively. Changing the assumed power-law index does not greatly affect the completeness and false discovery rate, which is expected since the high prevalence sources do not depend greatly on the assumed power-law index.

\begin{figure}
\plotone{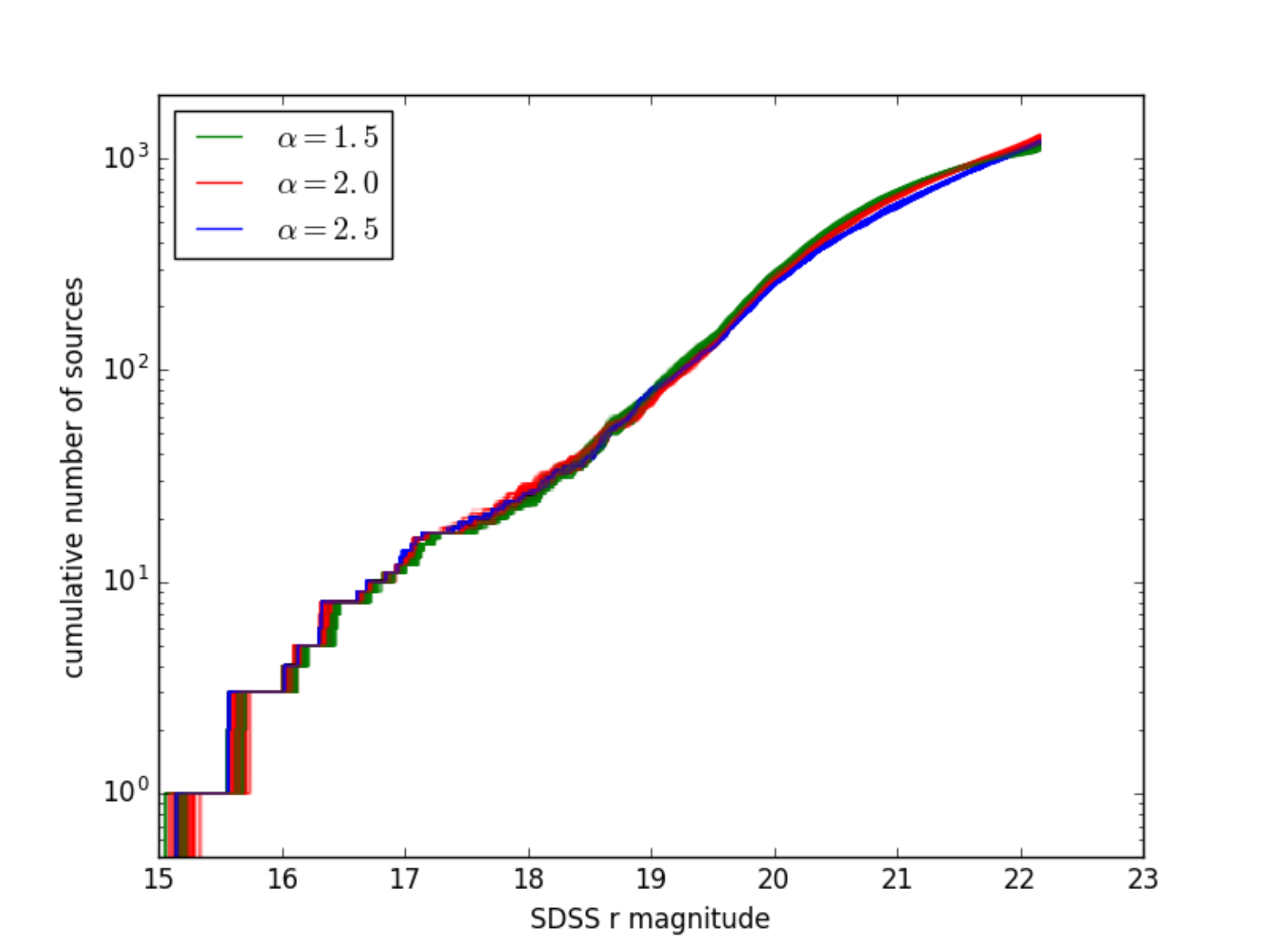}
\caption{Cumulative flux distributions of samples from the catalog ensemble, assuming different flux distribution power-law indices.}
    \label{fig:alpha_dist}
\end{figure}

\begin{figure}
\plotone{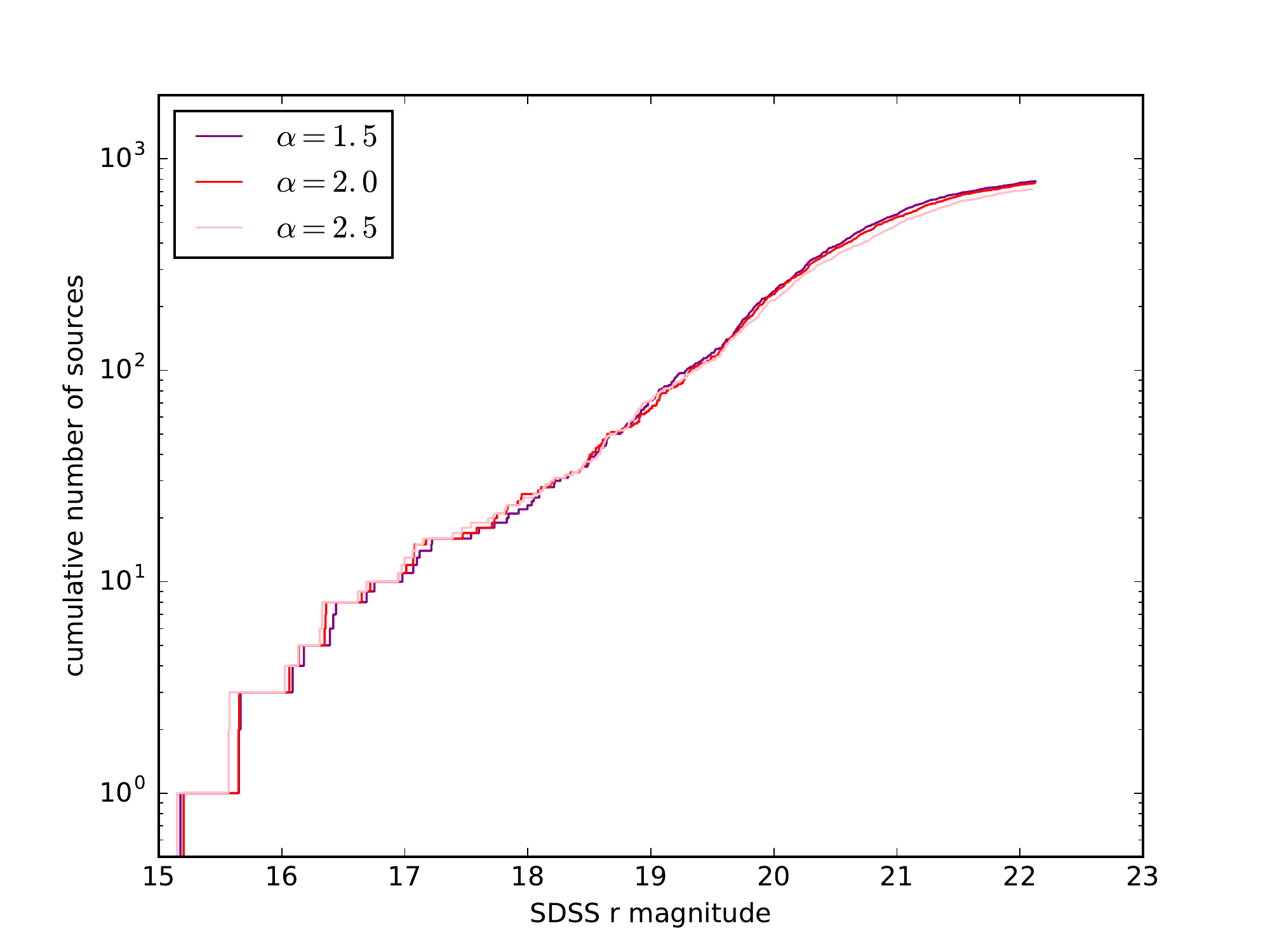}
\caption{Cumulative flux distributions of condensed catalog sources with over 95\% prevalence, assuming different flux distribution power-law indices.}
    \label{fig:alpha_prevalence}
\end{figure}

\begin{figure}
\plotone{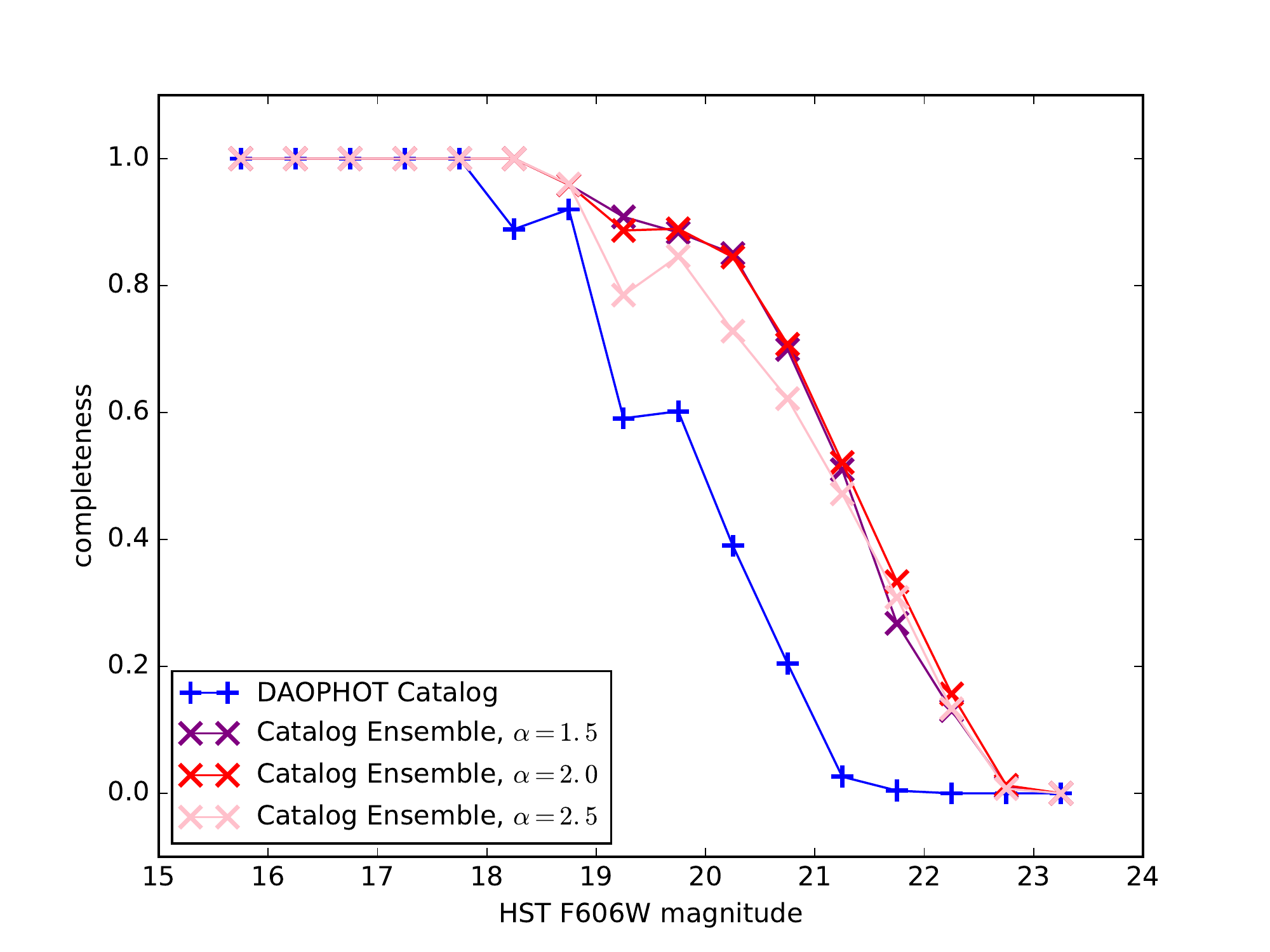}
\caption{Completeness of the catalog ensemble, assuming different flux distribution power-law indices, and the DAOPHOT catalog.}
    \label{fig:completeness_alpha}
\end{figure}

\begin{figure}
\plotone{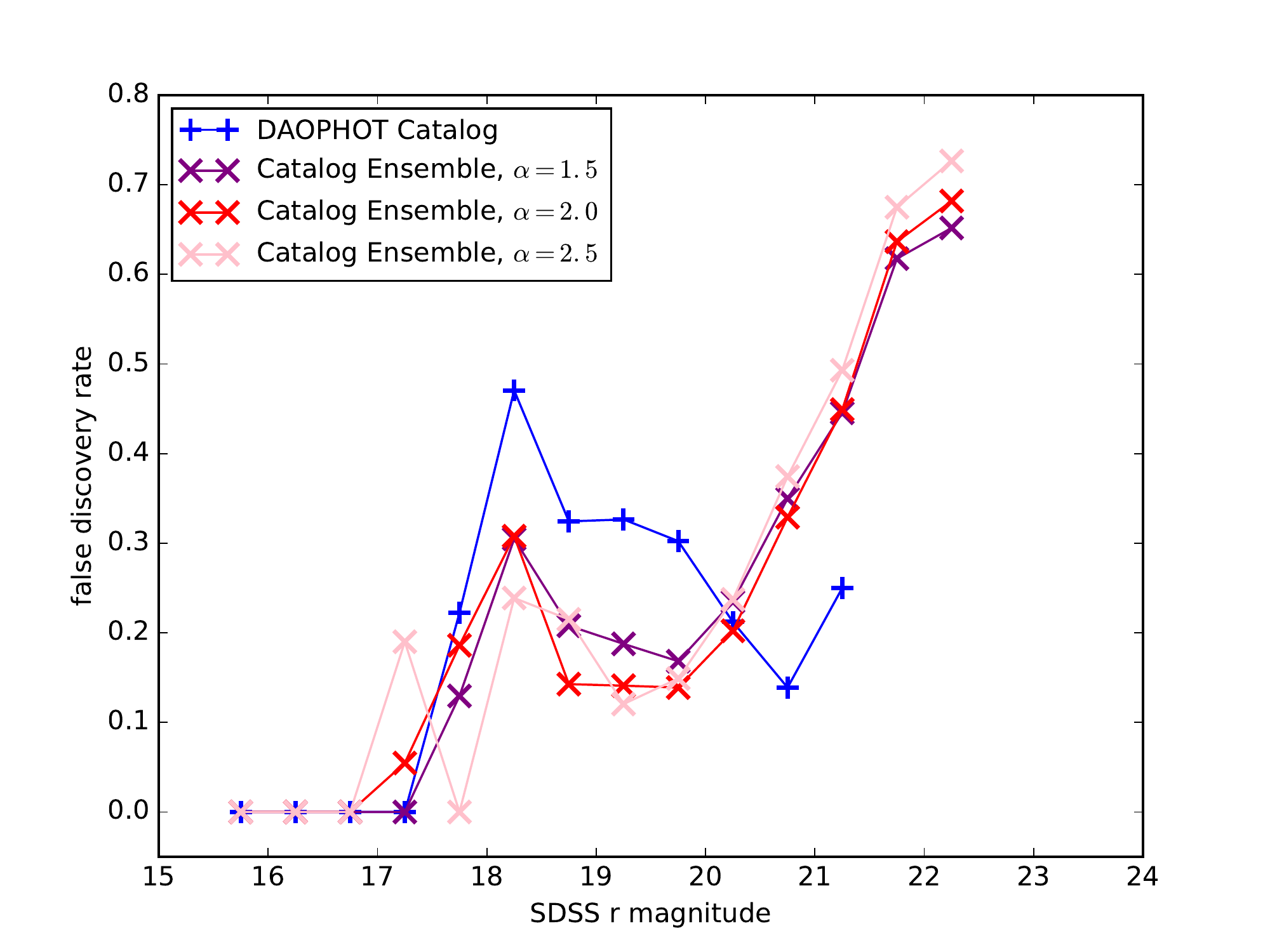}
\caption{False discovery rate of the catalog ensemble, assuming different flux distribution power-law indices, and the DAOPHOT catalog.}
    \label{fig:fdr_alpha}
\end{figure}
}

\added{
\section{Sensitivity to Assumed Point Spread Function}
\label{sec:psf_change}
We assume the PSF is known and take it to be the PSF determined by the SDSS pipeline. We make this choice because fitting for PSF parameters requires proposals which change all the sources in an image, making model image evaluation computationally expensive. To investigate the sensitivity of probabilistic cataloging to the assumed PSF, we take the SDSS PSF and perturb it by making it 10\% narrower and broader. We then re-catalog the image (which has not been changed) with these perturbed PSFs.

The perturbed PSFs have a drastic effect on the performance of the cataloger. The flux distributions of the catalog ensemble sources using the SDSS PSF and the perturbed PSFs are plotted in Figure \ref{fig:psf_dist}. When the PSF is too narrow, sources get oversplit, suppressing the number of bright sources and increasing the number of dim sources inferred. When the PSF is too wide, more blending occurs, increasing the inferred brightnesses of the brightest sources. These differences between the catalog ensembles can be seen in four different regions depicted in Figure \ref{fig:psf_panel}. The completenesses of these catalog ensembles is plotted in Figure \ref{fig:completeness_psf}. Using the narrow PSF decreases completeness at the bright end because these sources are bright enough that many sources are placed in the wings of their true PSF, decreasing the flux inferred for the source so much that it is no longer a match. Using a broad PSF makes the cataloger less sensitive to sources dimmer than 19\textsuperscript{th} magnitude. The false discovery rates of these catalog ensembles is plotted in Figure \ref{fig:fdr_psf}. The false discovery rate is worse for the perturbed PSFs than for the SDSS PSF, except that the broad PSF catalog ensemble has a lower false discovery rate than the SDSS PSF catalog ensemble between 17\textsuperscript{th} and 19\textsuperscript{th} magnitudes.

\begin{figure}
\plotone{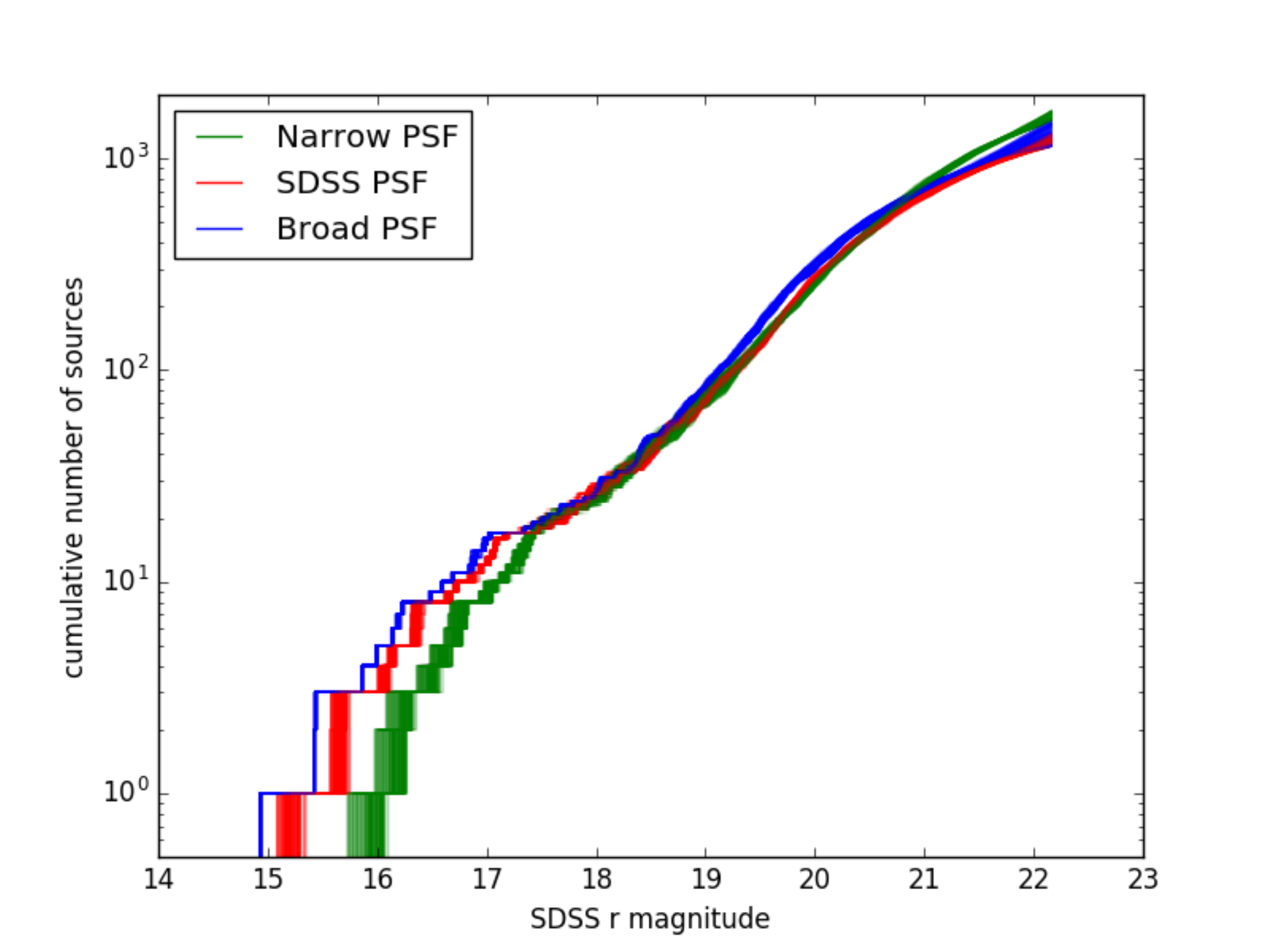}
\caption{Cumulative flux distributions of samples from the catalog ensemble, assuming different PSFs.}
    \label{fig:psf_dist}
\end{figure}

\begin{figure}
\plotone{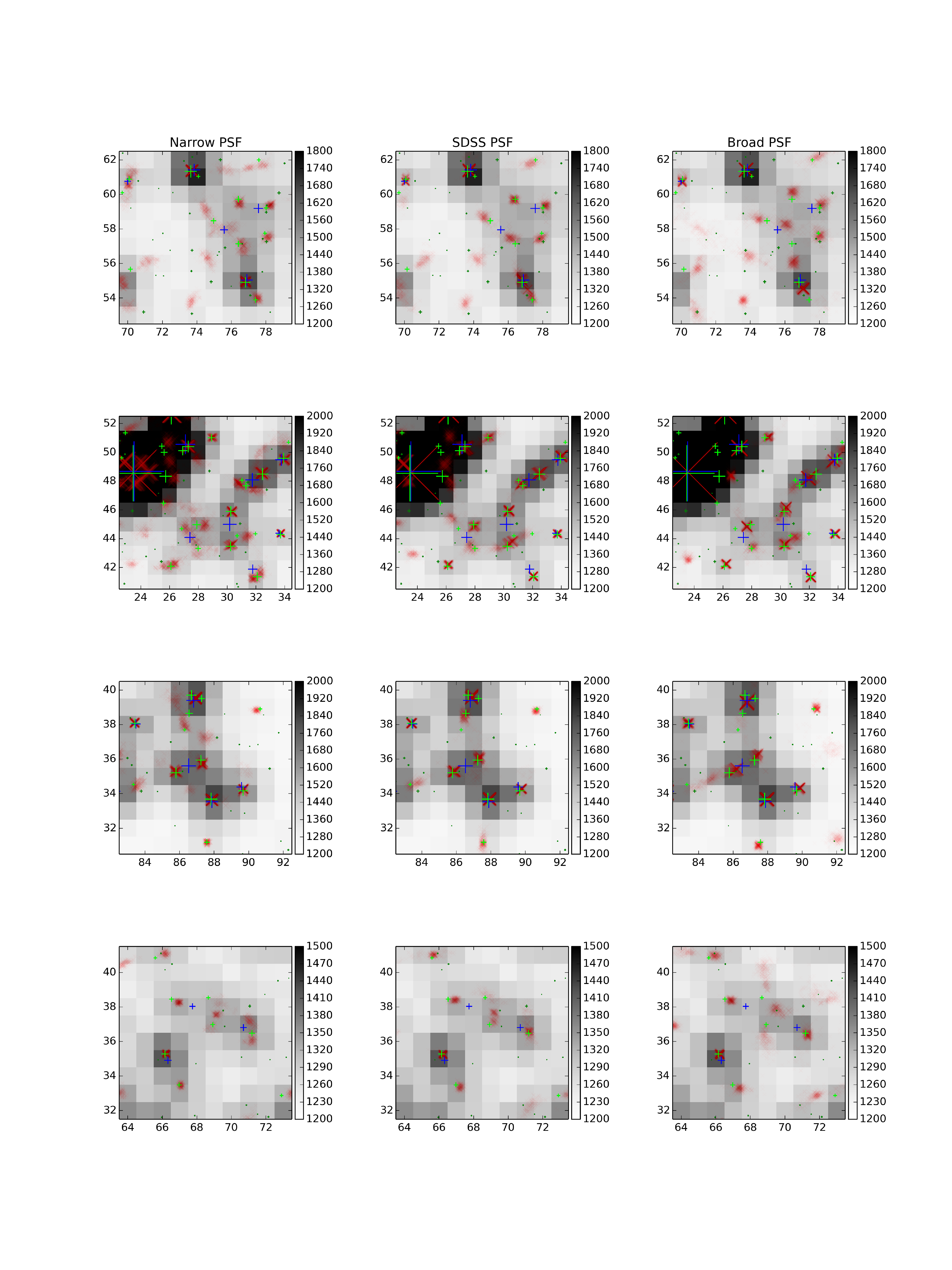}
\caption{Depiction of the catalog ensemble in four 10x10 pixel cutouts of the SDSS image, with different assumed PSFs. Each row of panels is a different region of the image, each column of panels has a different assumed PSF. Lime green crosses are HST sources brighter than 22\textsuperscript{nd} magnitude in F606W and dark green crosses are HST sources dimmer than this magnitude, both with area proportional to F606W flux. Blue crosses are DAOPHOT sources and translucent red Xs are sources from the catalog ensemble, stacked. The area of the DAOPHOT catalog and catalog ensemble source symbols is proportional to SDSS r band flux.}
    \label{fig:psf_panel}
\end{figure}

\begin{figure}
\plotone{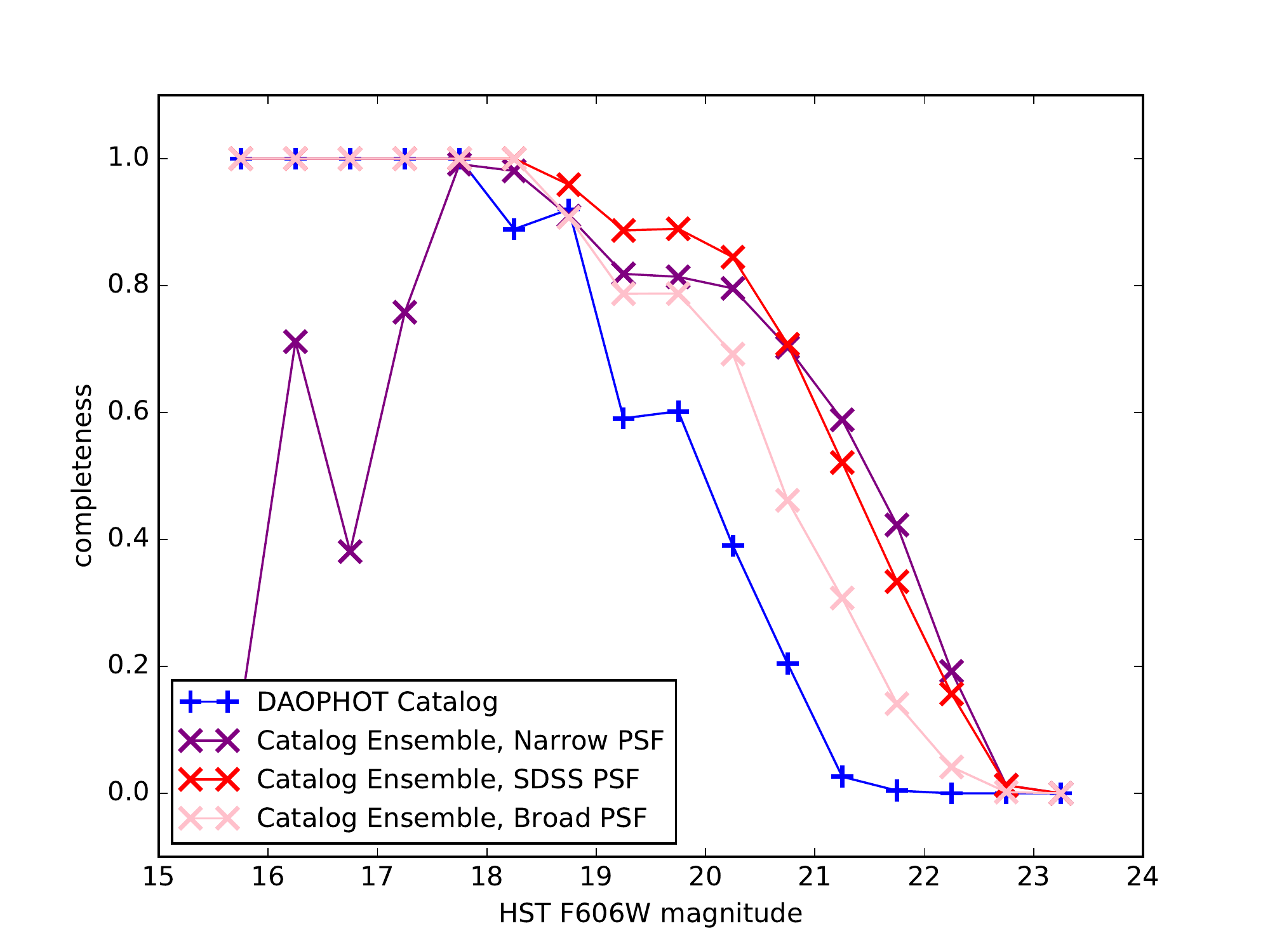}
\caption{Completeness of the catalog ensemble, assuming different PSFs, and the DAOPHOT catalog.}
    \label{fig:completeness_psf}
\end{figure}

\begin{figure}
\plotone{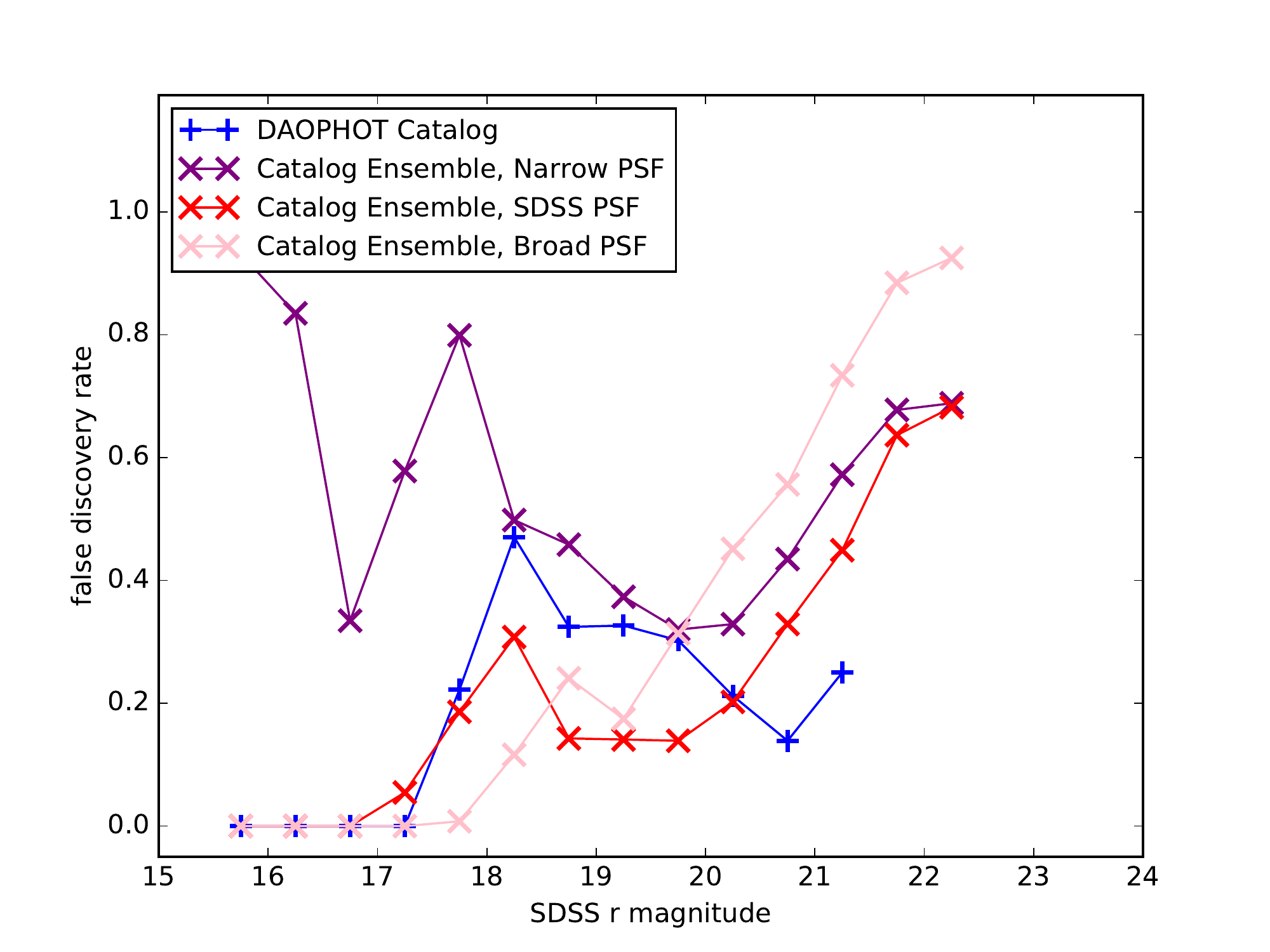}
\caption{False discovery rate of the catalog ensemble, assuming different PSFs, and the DAOPHOT catalog.}
    \label{fig:fdr_psf}
\end{figure}
}

\end{document}